\documentclass[11pt,a4paper]{article}
\pdfoutput=1
\usepackage{jheppub}

\usepackage{multirow, graphicx,amssymb,url,mathrsfs,amsmath}
\usepackage{wrapfig,boxedminipage,epsfig}
\usepackage{amsxtra,amstext,latexsym,dsfont,amsfonts}
\usepackage{color}
\usepackage[dvipsnames]{xcolor}
\usepackage{float}
\usepackage{slashed}
\usepackage{calligra}
\DeclareFontShape{T1}{calligra}{m}{n}{<->s*[2.2]callig15}{}
\DeclareMathAlphabet{\mathcalligra}{T1}{calligra}{m}{n}
\usepackage{tikz} 
\usepackage[compat=1.1.0]{tikz-feynman}
\usepackage{feynmf}
\usepackage[title]{appendix}
\usepackage{ulem}
\usepackage{contour}
\usepackage{subfigure}

\renewcommand{\l}{\lambda}

\newcommand{\be}{\begin{equation}}
\newcommand{\ee}{\end{equation}}
\newcommand{\bea}{\begin{eqnarray}}
\newcommand{\eea}{\end{eqnarray}}

\newcommand{\dd}{\mathrm{d}}

\usepackage{xparse}
\ExplSyntaxOn
\NewDocumentCommand{\HS}{m}
 {
  \seq_set_split:Nnn \l_tmpa_seq { ~ } { #1 }
  \seq_map_inline:Nn \l_tmpa_seq { \contour{green}{##1} ~ } \unskip
 }
\ExplSyntaxOff

\title{Holographic reconstruction of black hole spacetime: machine learning and entanglement entropy}

\author[a]{Byoungjoon Ahn,}
\author[b,c]{Hyun-Sik Jeong,}
\author[a,d]{Keun-Young Kim}
\author[a]{and Kwan Yun}

\affiliation[a]{Department of Physics and Photon Science, Gwangju Institute of Science and Technology, \\
123 Cheomdan-gwagiro, Gwangju 61005, Korea}
\affiliation[b]{Instituto de F\'isica Te\'orica UAM/CSIC, Calle Nicol\'as Cabrera 13-15, 28049 Madrid, Spain}
\affiliation[c]{Departamento de F\'isica Te\'orica, Universidad Aut{\'o}noma de Madrid, 28049 Madrid, Spain}
\affiliation[d]{Research Center for Photon Science Technology, Gwangju Institute of Science and Technology, \\
123 Cheomdan-gwagiro, Gwangju 61005, Korea}

\emailAdd{bjahn123@gist.ac.kr}
\emailAdd{hyunsik.jeong@csic.es}
\emailAdd{fortoe@gist.ac.kr}
\emailAdd{ludibriphy70@gm.gist.ac.kr}

\preprint{IFT-UAM/CSIC-24-88}

\abstract{
We investigate the bulk reconstruction of AdS black hole spacetime emergent from quantum entanglement within a machine learning framework. Utilizing neural ordinary differential equations alongside Monte-Carlo integration, we develop a method tailored for continuous training functions to extract the general isotropic bulk metric from entanglement entropy data. To validate our approach, we first apply our machine learning algorithm to holographic entanglement entropy data derived from the Gubser-Rocha and superconductor models, which serve as representative models of strongly coupled matters in holography. Our algorithm successfully extracts the corresponding bulk metrics from these data. Additionally, we extend our methodology to many-body systems by employing entanglement entropy data from a fermionic tight-binding chain at half filling, exemplifying critical one-dimensional systems, and derive the associated bulk metric. We find that the metrics for a tight-binding chain and the Gubser-Rocha model are similar. We speculate this similarity is due to the metallic property of these models.
}

\begin{document}
\maketitle

%%%%%%%%%%%%%%%%%%%%%%%%%%%
%
%%%%%%%%%%%%%%%%%%%%%%%%%%%
\section{Introduction}

Entanglement, a fundamental characteristic of quantum mechanics, elucidates the intriguing non-local correlations between quantum entities, and is at the core of quantum information sciences~\cite{Horodecki:2009aa,Aspect_1999,Nielsen_2012}. Especially, the notion of \textit{entanglement entropy} now exerts a wide-ranging influence, spanning from condensed matter~\cite{Amico:2008aa} to high-energy quantum field theory~\cite{Calabrese:2009qy} and even extending into quantum gravity~\cite{Nishioka:2009un}.

Within condensed matter physics, for example, entanglement entropy proves its versatility as a tool, facilitating the characterization of quantum phases and the intricate dynamics of strongly correlated many-body systems~\cite{Islam_2015,Laflorencie:2015eck}. The scaling behavior of entanglement entropy~\cite{Eisert:2010aa} offers insights into phases beyond symmetry characterization, particularly useful for identifying exotic states like topological phases~\cite{Kitaev:2006aa,Levin:2006zz,Jiang:2012uea} and spin liquids~\cite{Zhang:2011wm,Isakov:2011aa}. Additionally, entanglement entropy aids in exploring quantum criticality~\cite{Vidal:2002rm}, understanding non-equilibrium dynamics~\cite{Bardarson:2012fft,Daley:2012xhf}, and assessing numerical techniques for efficient many-body physics simulation~\cite{Schuch:2008zza}.

\paragraph{Holographic entanglement entropy.}
Nevertheless, the concept of entanglement entropy has also garnered increased attention in recent decades due to its significant role in holography (AdS/CFT correspondence)~\cite{Witten:1998zw,Gubser:1998bc,Maldacena1999}. Holography entails a duality between boundary quantum field theories and bulk gravitational theories, offering a compelling framework for understanding the emergence of spacetime geometry from quantum entanglement. Particularly notable is its connection exemplified in the Ryu-Takayanagi (RT) formula~\cite{Ryu:2006bv}, which establishes a relationship between the entanglement entropy of a subsystem in the boundary field theory and the area of the minimal surface in the bulk corresponding to the same region.\footnote{The RT formula, along with its subsequent generalizations~\cite{Hubeny:2007xt,Faulkner:2013ana,Dong2014,Dong:2016fnf}, provides a deeper elucidation of the pivotal role of entanglement in holographic duality. Another noteworthy entanglement measure is the reflected entropy, which possesses a gravity dual known as the entanglement wedge cross section~\cite{Takayanagi:2017knl,Nguyen:2017yqw,Dutta:2019gen,Jeong:2019xdr}.}

By leveraging the RT formula and understanding gravity through the lens of quantum entanglement, the holographic examination of entanglement entropy not only offers a more manageable approach to studying and computing this measure, but also yields general lessons applicable other branches of physics such, as quantum field theories and strongly coupled many-body physics~\cite{Rangamani_2017}. 
For instance, the $c$-theorem, established in two-dimensional conformal field theory, is derived from analyzing how entanglement entropy evolves under renormalization group flows~\cite{Zamolodchikov:1986gt}. Also, holographic entanglement entropy proves instrumental in the study of the aforementioned condensed matter physics by characterizing various phases of matter, exemplified by holographic superconductors~\cite{Albash:2012pd}.\footnote{Please refer to \cite{Jeong:2022zea} for a comprehensive list of additional references regarding the investigation of entanglement in diverse holographic matters.} This elucidates phenomena such as phase transitions, quantum criticality, and topological order, offering valuable insights into the entanglement properties of strongly correlated quantum systems.

\paragraph{Bulk reconstruction in holography.}
It is worth noting that, in most studies of holographic entanglement entropy, the standard approach is the bottom-up methodology. In this framework, gravitational bulk theories are employed to depict the realistic dual boundary quantum systems (e.g., characterized by broken translational symmetries), and the RT formula is used to study the associated entanglement entropy.

Nonetheless, despite significant advancements over the past decade, which provide compelling evidence that the entanglement structure of the underlying quantum mechanical degrees of freedom plays a pivotal role in shaping the emergent holographic spacetime geometry and its dynamics, an open question remains. Specifically, it is still unclear which  bulk gravity model can accurately reproduces the entanglement features of a given quantum system on the boundary and how to identify such models.

One essential approach for identifying such bulk gravity theories is termed `bulk reconstruction', representing a non-trivial inverse problem entailing holographic mapping from lower (boundary) to higher (bulk) dimensions. A captivating aspect within this endeavor is \textit{the reconstruction of the metric} in the holographic spacetime.\footnote{Additional significant branches can be found in~\cite{Hamilton:2006az,DeJonckheere:2017qkk,Harlow:2018fse,Kajuri:2020vxf}, with particular emphasis on bulk operator reconstruction.}
There are numerous methodologies for bulk metric reconstruction, leveraging diverse boundary physical quantities. These include the source and expectation value of the energy-momentum tensor~\cite{deHaro:2000vlm}, singularities in sets of correlation functions~\cite{Hammersley:2006cp,Hubeny:2006yu}, entanglement entropy of boundary intervals~\cite{Hammersley:2007ab,Bilson:2008ab,Bilson:2010ff,Hubeny:2012ry},\footnote{See also the recent work~\cite{Jokela:2023rba} where the holographic metric is constructed from the derivative of entanglement entropy in lattice Yang-Mills theory.} differential entropy, a UV-finite combination of entanglement entropy~\cite{Balasubramanian:2013lsa,Myers:2014jia,Czech:2014ppa}, divergence structure of boundary $n$-point functions~\cite{Engelhardt:2016wgb,Engelhardt:2016crc}, modular Hamiltonians of boundary subregions~\cite{Roy:2018ehv,Kabat:2018smf}, Wilson loops related to quark potential~\cite{Hashimoto:2020mrx}, four-point correlators in an excited quantum state~\cite{Caron-Huot:2022lff}, and holographic complexities~\cite{Hashimoto:2021umd,Xu:2023eof}, among others.
It is noteworthy that much of the research on metric reconstruction has been motivated by the notion that spacetime is constructed from quantum entanglement~\cite{Maldacena:2001kr,Ryu:2006bv,VanRaamsdonk:2010pw,Swingle:2009bg,Maldacena:2013xja,Ryu:2006ef}.\footnote{Another notable progression in investigating bulk reconstruction in holography involves a novel approach from quantum many-body systems, termed the multi-scale entanglement renormalization ansatz (MERA) of tensor networks~\cite{Vidal:2008zz,Swingle:2009bg,Vidal:2007hda,Milsted:2018san}.}

\paragraph{Machine learning and holography.}
In recent years, \textit{machine learning} has emerged as a promising approach to addressing bulk reconstruction by identifying the underlying bulk theory, such as reconstructing the bulk metric. Efficient holographic modeling has been successfully demonstrated through the application of machine learning techniques with various physical quantities, including lattice QCD data of the chiral condensate, hadron spectra, meson spectrum, shear viscosity, optical conductivity, and entanglement entropy, and so forth~\cite{Hashimoto:2018ftp, Hashimoto:2018bnb, Hashimoto:2019bih, Tan:2019czc, Akutagawa:2020yeo, Yan:2020wcd, Hashimoto:2020jug, Hashimoto:2021ihd, Katsube:2022ofz, Hashimoto:2022eij, Li:2022zjc, Park:2022fqy, Park:2023slm, Zhou:2023pti, Ahn:2024gjf, Gu:2024lrz, Chen:2024ckb, Hashimoto:2024rms, Chen:2024mmd, Bea:2024xgv, Mansouri:2024uwc}.\footnote{Other related works that extract the spacetime metric using machine learning, independent of holography, can be found in~\cite{You:2017guh,Hu:2019nea, Han:2019wue,Lam:2021ugb}.}

Machine learning holography not only facilitates the construction of data-driven holographic models but also provides insights into deeper understanding of holography itself. This approach is referred to as the AdS/DL correspondence, and establishes connections between deep learning (DL) -- a form of machine learning utilizing deep neural networks -- and holography. Here, deep learning functions as a solver for the inverse problem, enabling the determination of the bulk metric upon traning the neural network. Readers interested in grasping the essential concept of AdS/DL within a simplified framework are encouraged to refer \cite{Song:2020agw}, which illustrates this concept through a classical mechanics problem.

\paragraph{Motivation of this paper.}
In this manuscript, we investigate the reconstruction of the holographic bulk metric through machine learning of the boundary entanglement entropy. Our aim is to assess the effectiveness of machine learning in this context.

Note that a machine learning method has been applied in related contexts~\cite{Park:2022fqy, Park:2023slm}, where authors inferred the dual geometry from a given entanglement entropy using the RT formula and deep learning methods. However, due to inherent discontinuities in their algorithm, the reconstructed metric lacks continuity. Additionally, for simplicity, the entanglement entropy was assumed to have the dual geometry when only the blackening factor $f(z)$ is unknown, for instance, $h(z)=1$ in \eqref{OURMET}. We intend to enhance this work by incorporating a {neural ODE approach} and a more generic metric setup, i.e., any function of $f(z)$ and $h(z)$,\footnote{In the scenario of isotropic geometries, \eqref{OURMET} represents the most general metric setup. Specifically, metrics where ($g_{tt} \neq g_{zz} \neq g_{xx}$) can always be expressed in the form of \eqref{OURMET} through coordinate transformations. For detailed discussions, refer to \cite{Arean:2024pzo}.} making it more compatible with bulk reconstruction programs for continuous and generic metrics.

It is also worth mentioning that when $h(z)=1$, one may not need to resort to machine learning holography; instead, Bilson's method~\cite{Bilson:2010ff} can be utilized to invert the RT formula, generating the metric as output and entanglement entropy as input. Nevertheless, for completeness, we review Bilson's method in the main text and discuss its limitations when $h(z)\neq1$, motivating the use of machine learning holography to find the complete metric from entanglement entropy.

As illustrative examples for our purpose, in this study, we initially employ two representative \textit{holographic} entanglement entropy models with $h(z)\neq1$: the Gubser-Rocha model~\cite{Gubser:2009qt} and the holographic superconductor model~\cite{Hartnoll:2008vx, Hartnoll:2008kx}, where the former has been particularly instrumental in examining the characteristics of strange metals.\footnote{It is noteworthy that the Gubser-Rocha model~\cite{Gubser:2009qt} can be derived through a top-down approach. Specifically, for $d=3$, it results from a consistent truncation of eleven-dimensional supergravity compactified on AdS$_4 \times S^7$~\cite{Gubser:2009qt}, while for $d=4$, it is derived from ten-dimensional type IIB String Theory as the near-horizon limit of D3-branes~\cite{Gubser:2009qt, Cvetic:1999xp}. See also \cite{Gubser:2009qm, Gauntlett:2009dn} for discussions on top-down holographic superconductors from string/M-theory. Consequently, our study can also be regarded as an exploration of such top-down models through the lens of machine learning.} Moreover, in addition to the aforementioned holographic matter models, we incorporate entanglement entropy data from quantum many-body systems. Specifically, we utilize data obtained from a fermionic tight-binding chain at half filling in critical one-dimensional systems~\cite{Miao:2021aa} to ascertain the corresponding emergent bulk metric.

In essence, we propose applying data-driven and machine learning methodologies to unveil the emergent continuous metric based on boundary entanglement entropy data. 
Analyzing various entanglement entropy dataset within the \textit{generic and continuous} metric configuration, our objective is to establish machine learning holography as both a practical and theoretical framework. Through our investigation, we aim to offer valuable insights into the fundamental aspects of quantum entanglement, thereby deepening our understanding of many-body quantum physics and potentially advancing the exploration of holographic duality.

This paper is organized as follows. Section \ref{sec2} presents a quick review of the RT formula concerning the holographic entanglement entropy. Additionally, in section \ref{sec3}, we discuss Bilson's method for generating the metric from entanglement entropy and address its limitation in studying the complete metric. Section \ref{sec4} introduces a machine learning method tailored to determine the continuous/generic metric based on provided entanglement entropy. As input data of machine learning method, we utilize both the holographic entanglement entropy and entanglement entropy from a fermionic tight-binding chain. Finally, section \ref{sec5} is devoted to conclusions.

%%%%%%%%%%%%%%%%%%%%%%%%%%%
%
%%%%%%%%%%%%%%%%%%%%%%%%%%%
\section{Holographic entanglement entropy: a quick review}\label{sec2}
We consider the asymptotically AdS$_4$ spacetime as
\begin{align}\label{OURMET}
\begin{split}
\dd s^2 = \frac{L^2}{z^2} \left[ -f(z) \, \dd t^2 +  \frac{\dd z^2}{f(z)} + h(z) \, \left(\dd x^2 + \dd y^2\right) \right] \,,
\end{split}
\end{align}
where $L$ represents the AdS radius and the functions $f(z)$ and $h(z)$ are expanded as $f=h=1$ near the AdS boundary ($z\rightarrow0$).

For the later use, within the metric \eqref{OURMET} we also read the thermodynamic quantities as the temperature ($T$) and the thermal entropy ($s$) as
\begin{align}\label{TSfor}
\begin{split}
T = -\frac{f'(z)}{4\pi} \Bigg|_{z_{h}} \,, \qquad s = \frac{L^2}{4G_N} \frac{h(z)}{z^2} \Bigg|_{z_{h}} \,,
\end{split}
\end{align}
which are evaluated at the event horizon $z_h$, where $G_N$ is the Newton's constant.

\begin{figure}
\centering
\includegraphics[width=0.5\textwidth]{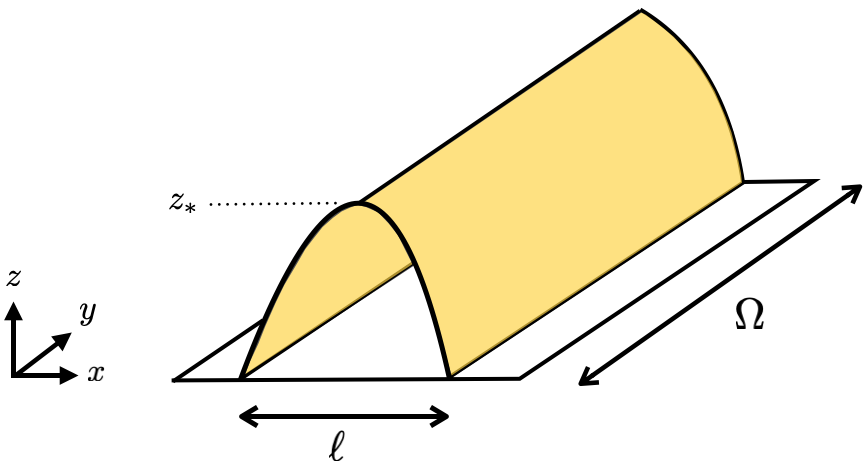}
\caption{A strip-shaped entangling region with its minimal surface depicted in yellow. The strip has a width $\ell$ in the $x$-direction and $\Omega$ in another direction. $z_{*}$ represents the maximum $z$ value attained by the minimal surface.}\label{sketchfig}
\end{figure}

\paragraph{Holographic entanglement entropy.}
The holographic dictionary to study the entanglement entropy ($S$) is the Ryu-Takayanagi formula~\cite{Ryu:2006bv,Ryu:2006ef}. The formula allows us to compute the entanglement entropy of the entangling region on the boundary by examining the corresponding region the bulk spacetime as
\begin{align}\label{EEFOR}
\begin{split}
S = \frac{\text{Area}(\gamma)}{4G_N} \,,
\end{split}
\end{align}
where $\gamma$ is the area if the minimal surface in the bulk, which is anchored at the AdS boundary. In this paper, we focus on a strip-shape entangling region. See Fig. \ref{sketchfig}: the minimal surface takes the form of a yellow surface with a width $\ell$ in $x$-direction and extends infinitely $\Omega$ in $y$-directions. $z_{*}$ represents the maximal $z$ value of this minimal surface in the bulk.

Within the metric \eqref{OURMET}, it is straightforward \cite{Ryu:2006bv,Ryu:2006ef} to find the geometric quantities for \eqref{EEFOR}, $S$ and $\ell$, as 
\begin{align}\label{SLfor}
\begin{split}
S = \frac{L^{2}\Omega}{2G_N} \int_{\epsilon}^{z_{*}} \, \frac{1}{z^{2}} \sqrt{\frac{h(z)}{1-\frac{z^{4}}{z_{*}^{4}} \frac{h(z_{*})^{2}}{h(z)^{2}} }} \frac{1}{\sqrt{f(z)}}\,\, \dd z  \,, \qquad
%%%
\ell = 2 \int_{0}^{z_{*}}  \frac{1}{\sqrt{\frac{h(\alpha)^{2}}{h(z_*)^{2}}\frac{z_{*}^{4}}{\alpha^{4}} - 1} } \frac{1}{\sqrt{h(\alpha)f(\alpha)}} \,\, \dd \alpha  \,,
\end{split}
\end{align}
where $\epsilon$ is the UV cutoff.
In order to isolate the UV divergence, it is convenience to split the integrand of $S$ in \eqref{SLfor} into two parts as 
\begin{align}\label{SFINITE}
\begin{split}
S &= \frac{L^{2}\Omega}{2G_N} \left[ \int_{\epsilon}^{z_{*}} \frac{1}{z^{2}} \dd z   + \int_{0}^{z_{*}} \, \frac{1}{z^{2}} \left(\sqrt{\frac{h(z)}{1-\frac{z^{4}}{z_{*}^{4}} \frac{h(z_{*})^{2}}{h(z)^{2}} }} \frac{1}{\sqrt{f(z)}} - 1 \right) \,\, \dd z   \right]  \\
&=  \frac{L^{2}\Omega}{2G_N} \left[  \frac{1}{\epsilon} + \left\{ \int_{0}^{z_{*}} \, \frac{1}{z^{2}} \left(\sqrt{\frac{h(z)}{1-\frac{z^{4}}{z_{*}^{4}} \frac{h(z_{*})^{2}}{h(z)^{2}} }} \frac{1}{\sqrt{f(z)}} - 1 \right) \,\, \dd z  - \frac{1}{z_{*}} \right\}  \right] \\
&:= \frac{L^{2}\Omega}{2G_N} \left[  \frac{1}{\epsilon} + S_{\text{Finite}}\right] \,,
\end{split}
\end{align}
where the UV divergence reads as $1/\epsilon$.

%%%%%%%%%%%%%%%%%%%%%%%%%%%
%
%%%%%%%%%%%%%%%%%%%%%%%%%%%
\section{Bilson's method}\label{sec3}

It is instructive to note that the Ryu-Takayanagi formula \eqref{SLfor} establishes a relationship where the entanglement entropy is derived from a given metric. In other words, the metric serves as the input data, while the entanglement entropy emerges as the output.

\subsection{Preliminary for the Bilson's method}
On the other hands, Bilson's method~\cite{Bilson:2010ff} attempts to invert the Ryu-Takayanagi formula \eqref{SLfor}, flipping the roles of metric and entanglement entropy. Here, the metric becomes the output, while the entanglement entropy becomes the input.

Bilson's method involves two main ingredients. Firstly, it employs the handbook of integral equations~\cite{polyanin1998handbook} to express:
\begin{align}\label{BMI}
\begin{split}
F(z_{*}) &= \int_{\epsilon}^{z_{*}}  \frac{1}{\sqrt{G(z_{*})-G(z)}} \, Y(z) \, \dd z  \,, \qquad G'(z_{*}) > 0 \\
(\text{Solution}): \quad Y(z_{*}) &= \frac{1}{\pi} \frac{\dd}{\dd z_{*}} \int_{\epsilon}^{z_{*}} \frac{G'(z)}{\sqrt{G(z_{*})-G(z)}} \, F(z) \, \dd z \,.
\end{split}
\end{align}
{The integral equation presented here shows how one function, $F(z_{*})$, can be expressed in terms of another, $Y(z_{*})$, with the roles reversed in the ``Solution" part of the equation.}
Another key aspect {of Bilson's method} involves {the connection between the entanglement entropy to the metric via the following derivative relationship}
\begin{align}\label{BMI2}
\begin{split}
\frac{\dd S}{\dd \ell}  = \frac{\dd z_{*}}{\dd \ell}  \frac{\dd S}{\dd z_{*}}  = \frac{L^{2} \Omega}{4 G_{N}} \frac{h(z_{*})}{z_{*}^{2}} \,,
\end{split}
\end{align}
where we used the chain rule {for derivatives, including how changes in the entanglement entropy $S$ with respect to the subsystem size $\ell$ are mediated through an intermediate variable $z_{*}$.}

{To understand \eqref{BMI2}, note that the chain rule splits the derivative into two parts: the changes of $S$ with respect to $z_{*}$, and the change of $z_{*}$ with respect to $\ell$. These components are further detailed as:}
\begin{align}\label{}
\begin{split}
\frac{\dd S}{\dd z_{*}} &=  \frac{L^{2} \Omega}{2 G_N}  \sqrt{h(z_{*})} \Bigg[ \lim_{z \rightarrow z_{*}}  \frac{1}{\sqrt{z_{*}^{4} - z^{4} \frac{h(z_{*})^{2}}{h(z)^{2}} } } \frac{1}{\sqrt{f(z_{*})}}  \\
& \qquad\qquad + z_{*}^{4} h(z_{*})^{2}  \int_{0}^{z_{*}} \sqrt{\frac{4 z^{10} h(z)^{6}  \left(2 h(z_{*}) - z_{*} h'(z_{*}) \right)^2 }{4 f(z) \left( z^2 z_{*}^{6} h(z)^3 h(z_{*}) - z^{6} z_{*}^{2} h(z) h(z_{*})^3  \right)^3}} \, \dd z \Bigg] \,,\\
%%%%%
\\
\frac{\dd \ell}{\dd z_{*}} &=   \frac{2 z_{*}^{2}}{\sqrt{h(z_{*})}} \Bigg[ \lim_{z \rightarrow z_{*}}  \frac{1}{\sqrt{z_{*}^{4} - z^{4} \frac{h(z_{*})^{2}}{h(z)^{2}} } }  \frac{1}{\sqrt{f(z_{*})}} \\
& \qquad\qquad + z_{*}^{4} h(z_{*})^{2}  \int_{0}^{z_{*}} \sqrt{\frac{4 z^{10} h(z)^{6} \left(2 h(z_{*}) - z_{*} h'(z_{*}) \right)^2 }{4 f(z) \left( z^2 z_{*}^{6} h(z)^3 h(z_{*}) - z^{6} z_{*}^{2} h(z) h(z_{*})^3  \right)^3}} \, \dd z \Bigg] \,.
\end{split}
\end{align}
{Here, we used a general relation for differentiating integrals:}
%Here, we also used the relation
%
\begin{align}\label{}
\begin{split}
U(z_{*}) = \int_{\epsilon}^{z_{*}}  H(z,\, z_{*}) \, \dd z    \quad \longrightarrow \quad
\frac{\dd U(z_{*})}{\dd z_{*}} =   H(z_{*},\, z_{*}) + \int_{\epsilon}^{z_{*}} \partial_{z_{*}} H(z,\, z_{*})  \, \dd z \,,
\end{split}
\end{align}
{where the first term involving the limit as $z$ approaches $z_{*}$ captures the local behavior near $z_{*}$, while the integral term encodes contributions from the entire interval $[0,\,z_{*}]$.}

%%%%%%%%%%%%%%%%%%%%%%%%%%%
\subsection{Reconstruction formula}

To leverage Bilson's method effectively, it is beneficial to choose the following coordinate
\begin{align}\label{}
\begin{split}
h(z) = 1   \,.
\end{split}
\end{align}
{This choice simplifies the mathematics significantly.}
We will address the general case where $h(z)\neq1$ towards the conclusion of this subsection.

\paragraph{Reconstruction formula I.}
Subsequently, we can recast the entanglement entropy \eqref{SLfor} into the form
\begin{align}\label{EE2}
\begin{split}
\frac{2G_N}{L^{2}\Omega} \, \frac{S(z_{*})}{z_{*}^{2}}  = \int_{\epsilon}^{z_{*}} \, \frac{1}{\sqrt{ z_{*}^{4} - z^{4} }} \frac{\sqrt{1/f(z)}}{z^{2}}  \,\, \dd z  \,.
\end{split}
\end{align}
{Comparing the expression \eqref{EE2} with the general formula \eqref{BMI}, we identify the functions $F(z_{*}),\,Y(z),\,\text{and }G(z)$ as}
\begin{align}\label{}
\begin{split}
F(z_{*}) := \frac{2G_N}{L^{2}\Omega} \, \frac{S(z_{*})}{z_{*}^{2}} \,, \qquad Y(z) := \frac{\sqrt{1/f(z)}}{z^{2}}  \,, \qquad
G(z) := z^{4} \,.
\end{split}
\end{align}
{Using these identification, the ``Solution" in \eqref{BMI} provides the reconstruction formula for $f(z)$ as}
\begin{align}\label{BMI3}
\begin{split}
\sqrt{\frac{1}{f(z)}} = \frac{8G_N}{\pi L^{2}\Omega} \,z^{2}\,  \frac{\dd}{\dd z} \int_{\epsilon}^{z} \, \frac{z_{*}}{\sqrt{ z^{4} - z_{*}^{4} }}  \,\, S(z_{*})   \,\, \dd z_{*}  \,,
\end{split}
\end{align}
together with \eqref{BMI2} as
\begin{align}\label{BMI4}
\begin{split}
\frac{\dd S (\ell)}{\dd \ell}   = \frac{L^{2} \Omega}{4 G_{N}} \frac{1}{z_{*}^{2}} \,.
\end{split}
\end{align}

The resulting expression \eqref{BMI3} represents the reconstruction formula provided by Bilson~\cite{Bilson:2010ff}, where $S(z_{*})$ is derived from the data $S(\ell)$, with $\ell$ interchangeable with $z_{*}$ through \eqref{BMI4}. Essentially, given the entanglement entropy $S(\ell)$, the metric component $f(z)$ can be generated as the output.

\paragraph{Reconstruction formula II.}
Although \eqref{BMI3} presents a novel approach, the entanglement entropy data $S(\ell)$ predominantly hinges on the UV cutoff $\epsilon$. Hence, from a practical standpoint, it proves beneficial \cite{Xu:2023eof} to employ an alternative reconstruction formula derived from the subsystem size in \eqref{SLfor}. Analogous to the entanglement entropy scenario, employing \eqref{BMI} yields:
\begin{align}\label{ECF2}
\begin{split}
\sqrt{\frac{1}{f(z)}} = \frac{2}{\pi} \frac{1}{z^{2}}  \frac{\dd}{\dd z} \int_{0}^{z} \, \frac{z_{*}^{3}}{\sqrt{ z^{4} - z_{*}^{4} }} \,\, \ell(z_{*})   \,\, \dd z_{*}  \,,
\end{split}
\end{align}
where $\ell(z_*)$ serves as the input and $f(z)$ emerges as the output.

However, it is noteworthy that the input data $\ell(z_*)$ essentially derives from \eqref{BMI4}, where the entanglement entropy $S(\ell)$ is provided. Thus, this second reconstruction formula \eqref{ECF2} also represents a scenario where the input data is the entanglement entropy.

\paragraph{Example: pure AdS geometry.}
{To better understand the application of the reconstruction formula,} we consider an example where the analytic expression for the entanglement entropy is available. For the pure AdS geometry by 
\begin{align}\label{PADSMET}
\begin{split}
f(z) = 1 \,,
\end{split}
\end{align}
the entanglement entropy \eqref{SLfor} can be analytically obtained as
\begin{align}\label{SAdS}
\begin{split}
S = S_{\text{CFT}} = \frac{L^{2}\Omega}{2G_N}  \left[ \frac{1}{\epsilon} - \frac{2\pi}{\ell} \left(\frac{\Gamma\left(\frac{3}{4}\right)}{\Gamma\left(\frac{1}{4}\right)}\right)^2 \right] \,.
\end{split}
\end{align}
As the initial step of Bilson's method, we can first put the given the entanglement entropy from \eqref{SAdS} into \eqref{BMI4} and find $\ell(z_{*})$
\begin{align}\label{}
\begin{split}
\ell(z_{*}) = \frac{2\sqrt{\pi} \, \Gamma\left(\frac{3}{4}\right)}{\Gamma\left(\frac{1}{4}\right)} \, z_{*} \,.
\end{split}
\end{align}
Then, by plugging this expression into the reconstruction formula \eqref{ECF2}, we can deduce
\begin{align}\label{}
\begin{split}
f(r) = 1 \,, 
\end{split}
\end{align}
{which is consistent with the expected pure AdS metric, as given in \eqref{PADSMET}.}

\paragraph{Limitation of Bilson's method.}
{While Bilson's method offers a structured approach to reconstructing the metric, it is not without its limitations.} It is straightforward to find that for the generic metric where
\begin{align}\label{}
\begin{split}
h(z) \neq 1   \,,
\end{split}
\end{align}
Bilson's reformulation formulas, \eqref{ECF2} and \eqref{BMI2}, become
\begin{align}\label{BMLIT}
\begin{split}
\sqrt{\frac{1}{f(z) \, h(z)^3}} &= \frac{2}{\pi} \frac{1}{z^{2}}  \frac{\dd}{\dd z} \int_{0}^{z} \, \frac{\ell(z_{*})}{\sqrt{ \frac{z^{4}}{h(z)^{2}} - \frac{z_{*}^{4}}{h(z_{*})^{2}} }} \,\,  \frac{2 z_{*}^{3} h(z_{*}) - z_{*}^{4} h'(z_{*}) }{2 h(z_{*})^3}  \,\, \dd z_{*}  \,, \\
%%%
\frac{\dd S}{\dd \ell}  &= \frac{L^{2} \Omega}{4 G_{N}} \frac{h(z_{*})}{z_{*}^{2}} \,.
\end{split}
\end{align}
Then, one can notice that the reconstruction program cannot proceed due to the inconsistency between the number of input data and output data. In other words, given a single data $S(\ell)$, one needs to determine two functions $f(z)$ and $h(z)$. Therefore, unless $h(z)$ is provided, Bilson's method cannot fully reconstruct the metric.

Nevertheless, this limitation is not surprising, as previously noted \cite{Bilson:2010ff}.
{An alternative approach that may alleviate some of these limitations involves transforming to a different coordinate system,}
%Interestingly, in an alternative coordinate defined by:
%
%
\begin{align}\label{ORICOR}
\begin{split}
\dd s^2 = \frac{L^2}{r^2} \left[ -g(r) e^{-\chi(r)} \dd t^2 + \frac{\dd r^2 }{g(r)} + \left(\dd x^2 + \dd y^2\right) \right]   \,,
\end{split}
\end{align}
{which simplifies the reconstruction formula}
% reconstruction formula is given by 
%
\begin{align}\label{ORI}
\begin{split}
\sqrt{\frac{1}{g(r)}} &= \frac{2}{\pi} \frac{1}{r^{2}}  \frac{\dd}{\dd r} \int_{0}^{r} \, \frac{r_{*}^{3}}{\sqrt{ r^{4} - r_{*}^{4} }} \,\, \ell(r_{*})   \,\, \dd r_{*}  \,, \\
%%%
\frac{\dd S}{\dd \ell}  &= \frac{L^{2} \Omega}{4 G_{N}} \frac{1}{r_{*}^{2}} \,.
\end{split}
\end{align}
Even though we can reconstruct $g(r)$ from the given $S(\ell)$ using \eqref{ORI}, {the inability to determine $\chi(r)$ means that the full metric cannot be completely specified.}\footnote{The metric \eqref{ORICOR} can be derived from our ansatz \eqref{OURMET} through the coordinate transformation:
\begin{align}\label{}
\begin{split}
r = \frac{z}{\sqrt{h(z)}} \,, \,\quad\, g(r) = \left(1-\frac{z h'(z)}{2h(z)}\right)^2  f(z)\,, \,\quad\, \chi(r) = \log \left[\left(1-\frac{z h'(z)}{2h(z)}\right)^2  h(z)\right] \,.
\end{split}
\end{align}
}\\

{In light of these challenges, the next section explores how machine learning techniques can be used to overcome the limitations of Bilson's method, enabling the reconstruction of both $f(z)$ and $h(z)$ from the given $S(\ell)$.}

%%%%%%%%%%%%%%%%%%%%%%%%%%%
\subsection{Example: Linear-axion model}\label{subsec33}

Despite the limitations that Bilson's method may entail, it remains a novel approach for constructing the blackening factor $f(z)$. We conclude this section by considering an example where bulk reconstruction via Bilson's method is viable: when $h(z)=1$.

In this context, we investigate the bulk reconstruction process when the entanglement entropy is provided numerically. To the best of our knowledge, our study represents the inaugural exploration of the numerical evaluation of the reconstruction formula \eqref{ECF2}.

\paragraph{Linear-axion model.}
For this purpose, we consider the linear-axion model~\cite{Andrade:2013gsa}, characterized by the action:
\begin{align}\label{LAMAC}
\begin{split}
    \mathcal{S} = \int \dd^4x \sqrt{-g} \left( R + 6 - \frac{1}{4} F^2 - \frac{1}{2}\sum_{I=1}^{2} (\partial \psi_I)^2 \right) \,,
\end{split}
\end{align}
where the field strength is $F=\dd A$. We set units $16\pi G_N = L = \Omega=1$ for simplicity hereafter.
Within the metric \eqref{OURMET}, the equations of motion from \eqref{LAMAC} allows the analytic background solutions as
\begin{align}\label{LAMMET}
\begin{split}
f(z) &= 1 - \frac{\beta^2}{2}z^2 -\left( 1 - \frac{\beta^2}{2} + \frac{\mu^2}{4} \right) z^3  + \frac{\mu^2}{4}z^4 \,, \\
A &= \mu \left(1 - z\right) \dd t \,, \qquad \psi_{I} = \beta (x\,,y) \,,
\end{split}
\end{align}
where the event horizon is set $z_h=1$, $\mu$ is the chemical potential, and $\beta$ the strength of broken translations.

It is noteworthy that axion theories~\cite{Andrade:2013gsa,Vegh:2013sk,Baggioli:2021xuv} have proven valuable in studying strongly coupled condensed matter systems~\cite{Zaanen:2015oix,Hartnoll:2016apf,Baggioli:2019rrs,Natsuume:2014sfa,Zaanen:2021llz,Faulkner:2010da} through analyses involving conductivity~\cite{Davison:2013txa,Gouteraux:2014hca,Blauvelt:2017koq,Jeong:2018tua,PhysRevLett.120.171602,Ammon:2019wci,Ahn:2019lrh,Jeong:2021wiu,Baggioli:2022pyb,Balm:2022bju,Ahn:2023ciq}, transport coefficients~\cite{Davison:2014lua,Blake:2016jnn,Amoretti:2016cad,Blake:2017qgd,Baggioli:2017ojd,Ahn:2017kvc,Giataganas:2017koz,Davison:2018ofp,Blake:2018leo,Jeong:2019zab,Arean:2020eus,Liu:2021qmt,Jeong:2021zhz,Wu:2021mkk,Jeong:2021zsv,Huh:2021ppg,Jeong:2022luo,Baggioli:2022uqb,Jeong:2023ynk,Ahn:2024aiw,Zhao:2023qms}, and the collective dynamics of strongly coupled phases~\cite{Baggioli:2014roa,Alberte:2015isw,Amoretti:2016bxs,Alberte:2017oqx,Amoretti:2017frz,Amoretti:2017axe,Alberte:2017cch,Andrade:2017cnc,Amoretti:2018tzw,Amoretti:2019cef,Ammon:2019wci,Baggioli:2019abx,Amoretti:2019kuf,Ammon:2019apj,Baggioli:2020edn,Amoretti:2021fch,Amoretti:2021lll,Wang:2021jfu,Zhong:2022mok,Bajec:2024jez}. Additionally, they have been explored in quantum information applications~\cite{RezaMohammadiMozaffar:2016lbo,Yekta:2020wup,Li:2019rpp,Zhou:2019xzc,Huang:2019zph,Jeong:2022zea,HosseiniMansoori:2022hok} and the application of the AdS/Deep learning correspondence~\cite{Ahn:2024gjf}.\footnote{Therefore, in this regard, our investigation into the linear-axion model \eqref{LAMAC} can be viewed as the bulk reconstruction of toy models for holographic strongly coupled materials within Bilson's method for the first time.}

It is worth noting that \cite{Ahn:2024gjf} successfully generated the metric of linear-axion models based on given boundary optical conductivity data using a machine learning method. Here, we aim to demonstrate that the reconstruction formula \eqref{ECF2} can yield the same bulk metric obtained from the machine learning method \cite{Ahn:2024gjf}.
To accomplish this, we utilize the same parameter set employed in \cite{Ahn:2024gjf}:
\begin{align}\label{LAMPARA}
\begin{split}
(\mu,\, \beta)=
\begin{cases}
\,\, (0.5,\, 0.5)\,, \quad\, (\text{Data 1})  \\
%%%
\,\, (0.5,\, 1.0)\,, \quad\, (\text{Data 2})  \\
%%%
\,\, (1.0,\, 1.5)\,. \quad\,\, (\text{Data 3}) 
\end{cases}
\end{split}
\end{align}

\paragraph{Holographic entanglement entropy.}
Plugging \eqref{LAMMET} into \eqref{SLfor}, one can numerically evaluate the entanglement entropy. Nevertheless, due to the UV divergence, it is advantageous to examine its finite piece, denoted as $S_{\text{Finite}}$, given in \eqref{SFINITE}.

\begin{figure}
\centering
\includegraphics[width=0.5\textwidth]{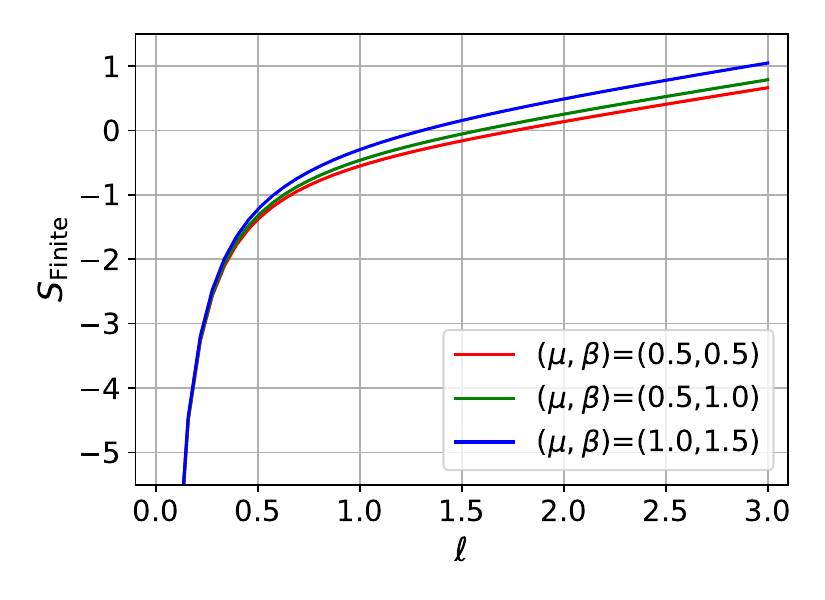}
\caption{The finite piece of the entanglement entropy of linear-axion models when $(\mu,\, \beta)=$ Data 1 (red), Data 2 (green), Data 3 (blue): see \eqref{LAMPARA}. The numerically obtained $S_{\text{Finite}}$ is aligned with the fitting \eqref{FITTING1} with the coefficients in Table. \ref{Table1}.}\label{RNFIGdata}
\end{figure}

Considering the parameters provided in \eqref{LAMPARA}, we illustrate $S_{\text{Finite}}$ in Fig. \ref{RNFIGdata}.\footnote{Strictly speaking, the corresponding entanglement entropy is $S_{\text{Finite}} \, z_h$ and $\ell/z_h$.}
Then, to utilize the numerically evaluated entanglement entropy $S_{\text{Finite}}$ as the input data for Bilson's method, we employ the power-law fitting curve
\begin{align}\label{FITTING1}
\begin{split}
S_{\text{Finite}} \,=\,   \sum_{i=-1}^{c_{\text{max}}} \,c_{i} \, \ell^{i}   \,,
\end{split}
\end{align}
where we set $c_{\text{max}}=5$ in this paper. For instance, we find that the numerically obtained $S_{\text{Finite}}$ is matched with the fitting coefficients given in Table. \ref{Table1}.
\begin{table}[h]
\centering
\begin{tabular}{lrrrrrrr}
             & $c_{\,-1}$ & $c_{0}$ & $c_{1}$ & $c_{2}$  & $c_{3}$ & $c_{4}$ & $c_{5}$   \\
\hline
\hline
Data 1 & -0.71778 & 0.00109 & 0.03538 & 0.15593 & -0.03102  & 0.00320 & -0.00012   \\
Data 2 & -0.71777 & 0.00041 & 0.17841 & 0.08822 & -0.01526 & 0.00151 & -0.00063    \\
Data 3 & -0.71775 & -0.00084 & 0.41694 & 0.00291 & -0.00039 & 0.00032 & -0.00002
\end{tabular}
\caption{Numerical fitting coefficients for \eqref{FITTING1} of the linear-axion model.}\label{Table1}
\end{table}
Note that the value of $c_{\,-1}$ originates from the finite piece of pure AdS geometry \eqref{SAdS}: 
\begin{align}\label{}
\begin{split}
- 2\pi \left(\frac{\Gamma\left(\frac{3}{4}\right)}{\Gamma\left(\frac{1}{4}\right)}\right)^2 \approx -0.71777   \,.
\end{split}
\end{align}
Additionally, $c_{0}\approx0$ aligns with the analytic examination of the small $\ell$ limit analysis of entanglement entropy for the linear-axion model, as provided in \cite{Jeong:2022zea}.

\begin{figure}
\centering
\includegraphics[width=0.47\textwidth]{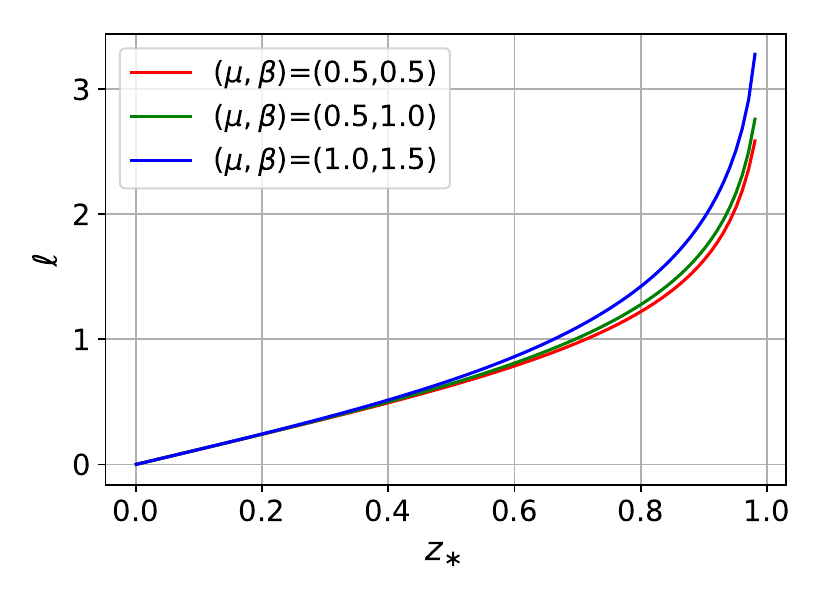}
\,
\includegraphics[width=0.48\textwidth]{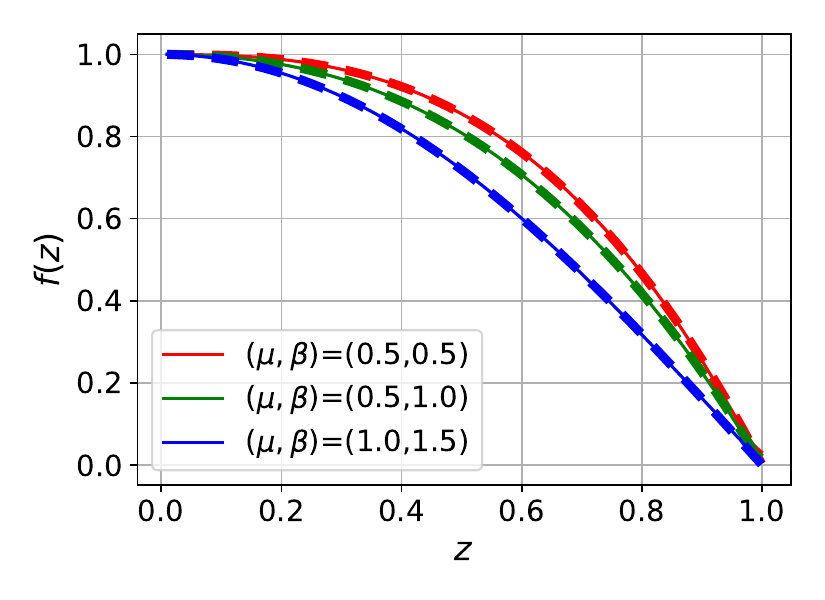}
\caption{The subsystem size $\ell(z_{*})$ (left panel) and the blackening factor $f(z)$ (right panel)  when $(\mu,\, \beta)=$ Data 1 (red), Data 2 (green), Data 3 (blue). The solid line is the one from Bilson's method \eqref{BMI4} or \eqref{ECF2}, while the dashed line is $f(z)$ in \eqref{LAMMET}.}\label{BMRfig1}
\end{figure}

\paragraph{Bilson's method and bulk reconstruction.}
Now that we have the expression of $S_{\text{Finite}}$ from \eqref{FITTING1}, we can determine the entanglement entropy $S$ using \eqref{SFINITE}. Subsequently, similar to the pure AdS example in \eqref{PADSMET}, the next step for Bilson's method entails inserting the numerically obtained $S$ into the reconstruction formula \eqref{BMI4} to determine $\ell(z_{*})$. However, in this instance, we numerically evaluate $\ell(z_{*})$ (as illustrated in the left panel of Fig. \ref{BMRfig1}) and subsequently determine the bulk metric component $f(z)$ numerically via the reconstruction formula \eqref{ECF2}.

The resulting $f(z)$ from Bilson's method is depicted by the solid line in the right panel of Fig. \ref{BMRfig1}. We confirm that this $f(z)$ successfully reconstructs the target metric $f(r)$ specified in \eqref{LAMMET}. Therefore, our findings demonstrate that the metric of the linear-axion model can be constructed not only via the machine learning method utilizing optical conductivity \cite{Ahn:2024gjf}, but also through Bilson's method \eqref{ECF2} with entanglement entropy.

%%%%%%%%%%%%%%%%%%%%%%%%%%%
%
%%%%%%%%%%%%%%%%%%%%%%%%%%%
\section{Machine learning method}\label{sec4}

In this section, we investigate bulk reconstruction employing the machine learning approach, which plays a pivotal role in revealing the complete metric described in \eqref{OURMET}. Our analysis demonstrates that by implementing machine learning techniques, we can overcome the constraints inherent in Bilson's method: specifically, we achieve the reconstruction of both $f(z)$ and $h(z)$ from the provided $S(\ell)$.

%%%%%%%%%%%%%%%%%%%%%%%%%%%
\subsection{Methodology: Neural ODEs and Monte-Carlo integration}

As discussed in the introduction, the machine learning holography framework, also known as the AdS/DL correspondence, has been employed in elucidating the holographic bulk theory that underpins quantum systems on the boundary.

In the pioneering studies of the AdS/DL correspondence~\cite{Hashimoto:2018ftp, Hashimoto:2018bnb}, significant advancements have primarily been achieved by \textit{discretizing} the holographic bulk metric using deep learning techniques. For example, the ordinary differential equation (ODE) can be solved within the framework of a residual neural network (ResNet)~\cite{E:2017}, which is composed of residual blocks~\cite{He:2016}.

In recent years, to achieve a \textit{continuous} holographic bulk spacetime, a continuous version of ResNet, known as the neural ordinary differential equation (neural ODE)~\cite{Chen:2018aa}, has been developed. By employing neural ODEs in machine learning holography, the continuous bulk metric has been reconstructed using boundary data, as demonstrated in the contexts of quantum chromodynamics~\cite{Hashimoto:2020jug} and strongly interacting condensed matter systems~\cite{Ahn:2024gjf, Gu:2024lrz}.

\paragraph{Neural ODEs.}
In this paper, we utilize the neural ODE framework within the context of the holographic study of quantum information, with a particular focus on entanglement entropy. Our machine learning methodology represents an integrated version of the neural ODE, which we refer to as neural integration.

First, let us briefly review the concept of the ResNet using a simple ODE example:
\begin{equation}\label{RESNETODEEQ}
\partial_z\mathcal{F} = \mathcal{G}(z,\mathcal{F};\theta(z)) \,,
\end{equation}
where $\mathcal{F}$ is the unknown function of $z$, $\mathcal{G}$ is a given function. Here, $\theta(z)$ represents a general hidden function within the ODE, which serves as the trainable parameters in the neural network.
This equation can be solved using ODE solvers by discretizing the $z$ interval into small steps $\Delta z$. For example, if we choose the Euler method as the solver, the approximate numerical solution $\mathcal{F}_N$ is given by
\begin{equation}\label{ResNetEM}
    \mathcal{F}_N = \mathcal{F}_1 + \sum_{n=1}^{N-1} \mathcal{G}(z_n,\mathcal{F}_n;\theta_n)\cdot\Delta z \,,
\end{equation}
where the $z$ interval is discretized into $N$ steps, and $\mathcal{F}_1$ is the given initial value at $z_1$ (for more details, see \cite{Ahn:2024gjf}). Due to this discretization, the training parameters are also discretized; for instance, $\theta_n(z_n)$ could be the discretized metric $f_n(z_n)$. This is the fundamental concept of ResNet for generating the discretized bulk metric~\cite{Hashimoto:2018ftp, Hashimoto:2018bnb}.

The approximate numerical solution $\mathcal{F}$ can be more generally expressed as:
\begin{equation}
    \mathcal{F}_{\textrm{final}} = \textrm{ODE\;Solver}\left[\mathcal{G};\, \mathcal{F}_{\textrm{initial}},\, (z_{\textrm{final}},\,z_{\textrm{initial}});\, \theta(z) \right] \,.
\end{equation}
For instance, for the Euler method described in \eqref{ResNetEM}, we have 
\begin{equation}
(\mathcal{F}_{\textrm{initial}},\,\mathcal{F}_{\textrm{final}}) = (\mathcal{F}_{1},\,\mathcal{F}_{N}) \quad \text{and}\,\quad (z_{\textrm{initial}},\,z_{\textrm{final}})=(1,\,N-1) \,,
\end{equation}
with the training parameter $\theta$ chosen as $\theta_n$ for the ResNet.\\

To ensure that the discrete hidden function becomes a continuous function, the central idea of neural ODEs~\cite{Chen:2018aa} is to replace the discrete parameter $\theta(z)=\theta_n(z_n)$ with a continuous \textit{deep neural network} $\theta(z) = D(z)$. This deep neural network consists of multiple layers through which data is processed, including an input layer, several hidden layers, and an output layer.

The deep neural network is defined as follows
\begin{equation}\label{DNN}
    D(z) = W_M\cdot \phi(\cdots\phi(W_2\cdot\phi(W_1z+b_1)+b_2)\cdots)+b_M \,,
\end{equation}
where $\phi$ is an activation function, $W_M$ is the weight matrix of $M$-th layer, and $b_M$ is the bias vector of $M$-th layer. The activation function is a non-linear function applied to the output of each layer, enabling the network to model complex patterns. In this paper, we use the rectified linear unit (ReLU) as the activation function: $\text{ReLU}(z)=\text{max}(0,z)$.
The other parameters (weights and biases) are adjusted during training to minimize errors in the loss function. 

\begin{figure}
    \centering
    \includegraphics*[viewport=0 0 750 340, width=0.8\textwidth]{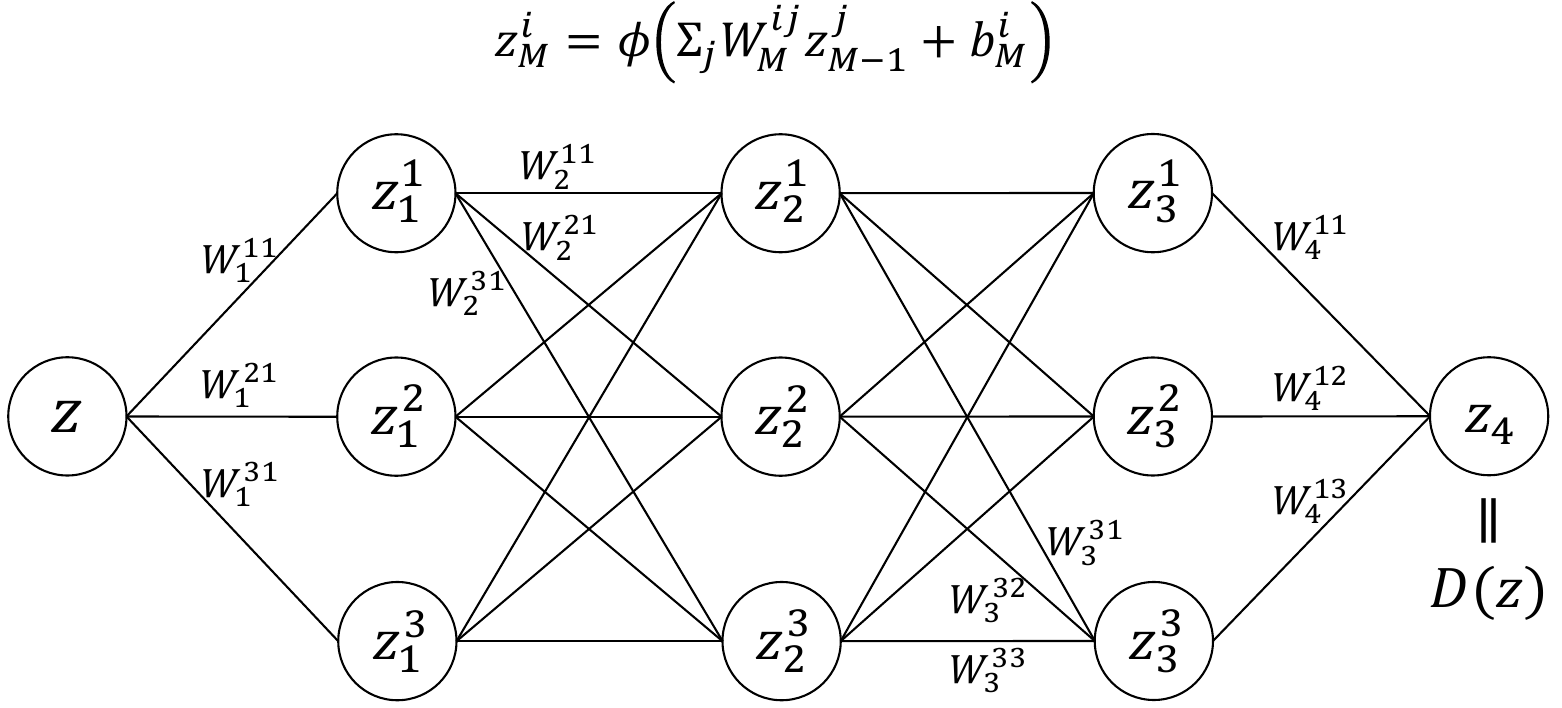}
    \caption{A structure of a deep neural network. It consists of 3 hidden layers with 3 nodes each. The nodes are fully connected each other. Through the learning procedure, the continuous function $D(z)$ is generated where the input is $z=:z_{0}^{1}$.}\label{fig:DNN}
\end{figure}

Continuous training variables in deep neural networks, primarily weights and biases, are essential for the learning process. Weights determine the ``strength" of connections between nodes, while biases provide each layer with a trainable constant value to help the model fit the data better. For an illustrative sketch of a deep neural network, see Fig. \ref{fig:DNN}, which shows the case of 3 hidden layers with 3 nodes each.
By employing iterative forward and backward propagation steps, along with optimization algorithms, these variables are continuously refined to minimize errors and enhance the model's efficacy.

In our deep neural network computations, we utilize the \textit{PyTorch} framework \cite{NEURIPS2019_9015}, an open-source machine learning library, and deep neural networks are initialized with 3 hidden layers, each containing 20 nodes.

\paragraph{Neural integrations.} 
In this manuscript, drawing inspiration from solving ODEs using deep neural networks, we introduce a method for solving integrations through machine learning techniques. Termed ``neural integration," this approach aligns well with our objective: employing machine learning holography to analyze entanglement entropy \eqref{SFINITE}.

To demonstrate further, let us consider the following integration, which can be obtained by equation \eqref{RESNETODEEQ}:
\begin{equation}\label{NEintegration}
    \mathcal{F} = \int_{z_\textrm{initial}}^{z_\textrm{final}} \mathcal{G}(z; D(z)) \dd z \,,
\end{equation}
where we substitute $\theta(z)$ with the deep neural network $D(z)$.
Subsequently, our primary objectives, the subsystem size $\ell$ \eqref{SLfor}, and the (finite piece) entanglement entropy $S_{\text{Finite}}$ \eqref{SFINITE}, can be expressed in the form of \eqref{NEintegration} as 
\begin{align}\label{SFINITE_neural integration}
\begin{split}
S_{\text{Finite}} =  \int_0^{z_{*}} \mathcal{G}_S(z; \, f(z),h(z)) \,\,\dd z  - \frac{1}{z_{*}} \,, \qquad
\ell = \int_0^{z_{*}} \mathcal{G}_\ell(z; \, f(z),h(z)) \,\,\dd z \,,
\end{split}
\end{align}
where $S_{\text{Finite}}$ (and $\ell$) is compared with $\mathcal{F}$, and the metric components with the deep neural network $D(z)$ in \eqref{DNN}, i.e.,
\begin{align}\label{metric to DNN}
\begin{split}
f(z) \,\,\longleftrightarrow\,\, D_f(z) \,, \qquad h(z) \,\,\longleftrightarrow\,\, D_h(z) \,.
\end{split}
\end{align}

The optimization of deep neural networks conducts through the learning process to achieve convergence between the estimated and intended values (given the input dataset $S_\text{Finite}$ and $\ell$). The estimated value is obtained via numerical integration in \eqref{SFINITE_neural integration}. In this study, Monte Carlo integration serves as the numerical integrator, with \textit{Torchquad} \cite{Gomez:2021czl}, an extension package of \textit{PyTorch}, utilized for Monte Carlo integration.\footnote{It is noteworthy that while bulk holographic geometry is derived from given entanglement entropy using deep learning methods \cite{Park:2022fqy, Park:2023slm}, the authors in \cite{Park:2022fqy} adopt ResNet together with the Runge-Kutta method within the ODE framework \eqref{RESNETODEEQ}, ensuring that the emergent metric remains a discrete function. Using the different machine learning method based on the transformer algorithm, the discrete metric is also generated in \cite{Park:2023slm}.}

\paragraph{More on the setup for machine learning holography.}
We further elaborate on the setup for our machine learning methodology. Instead of utilizing \eqref{metric to DNN}, we adopt the following technically convenient metric ansatz:
\begin{equation}\label{METANDNN}
    f(z) = (1-z)\left[1+(a+1)z-z^2 \, D_f(z)\right] \,, \qquad
    h(z) = 1 + az - z^2 \, D_h(z)   \,. 
\end{equation}
This choice is motivated by several reasons. Firstly, we impose the horizon condition at the horizon ($z=1$) as $f(1)=0$, and the asymptotic AdS boundary condition at the boundary ($z=0$) as $f(0)=h(0)=1$. Furthermore, from the near-boundary expansion of $\mathcal{G}_S(z; \, f(z),h(z))$ in \eqref{SFINITE_neural integration} together with \eqref{SFINITE}, it follows that 
\begin{align}\label{SFINITE_Series}
    \begin{split}
    \mathcal{G}_S(z; \, f(z),h(z)) \approx  \frac{h'(0) - f'(0)}{z}  \,,
    \end{split}
\end{align}
which would result in a logarithmic UV divergence in $S_{\text{Finite}}$, which is assumed to be absent in our study.\footnote{Within the models of interest, we have checked that such an additional UV divergence term is absent.} Within our ansatz \eqref{METANDNN}, we eliminate this UV divergence since $h'(0)=f'(0)=a$. Essentially, within our ansatz \eqref{METANDNN}, we have three training parameters in our machine learning method: the linear coefficient $a$, and the weights and biases of $D_f(z)$ and $D_h(z)$.

Then, substituting \eqref{METANDNN} into \eqref{SFINITE_neural integration} and initializing random values for $a$, weights, and biases, $S_{\text{Finite}}(\ell)$ can be computed. As an intermediate numerical step, both $S_{\text{Finite}}(z_{*})$ and $\ell(z_{*})$ are evaluated within the selected range $z_{*} \in [0.1,\,0.99]$.\\

Last but not least, we discuss the loss function $L$ employed in our computation, which is defined as
\begin{equation}\label{LossFUNC}
    L \,=\, \frac{1}{N_\ell}\sum_\ell \left| {S}^{(m)}_{\text{Finite}}(\ell) - {S_{\text{Finite}}}(\ell) \right| \,+\, \left|s^{(m)} - s\right| \,,
\end{equation}
where $N_{\ell}$ denotes the number of the data points for $\ell$. The first term quantifies the disparity between  the output entanglement entropy data ${S}^{(m)}_{\text{Finite}}(\ell)$ computed by the machine learning and the provided input data ${S_{\text{Finite}}}(\ell)$.
The second term, referred to as the penalty term, corresponds to the thermal entropy $(s)$ \eqref{TSfor}, where the superscript index ($m$) indicates that the thermal entropy is obtained through machine learning. This term is important for shaping the loss function and influencing the the optimization process.

In principle, alternative penalty terms could be chosen. However, opting for the thermal entropy as the penalty term offers several intriguing aspects.
Firstly, it aligns structurally with the first term in the loss function. In other words, $s^{(m)}$ is obtained through machine learning, while $s$ is given from the input data. It is worth recalling that thermal entropy can be inferred from the large subsystem size limit as $S_{\text{Finite}}\approx s \, \ell$~\cite{Ryu:2006bv,Ryu:2006ef}.
Furthermore, through the utilization of thermal entropy, an additional horizon condition on $h(1)$ can be introduced via \eqref{TSfor}, complementing the demonstration provided below \eqref{METANDNN}. Thus, in aggregate, we have imposed two AdS boundary conditions and two horizon condition on the metric components.

%%%%%%%%%%%%%%%%%%%%%%%%%%%
\subsection{Emergent spacetime from holographic entanglement entropy}

Utilizing the aforementioned machine learning methodologies, we commence with the training procedure. Initially, within this section, we illustrate the emergence of full spacetime from holographic entanglement entropy. To achieve this, we employ two exemplary and extensively researched toy models of holographic matters, namely the Gubser-Rocha model~\cite{Gubser:2009qt} and holographic superconductor model~\cite{Hartnoll:2008vx,Hartnoll:2008kx}, where the former has been instrumental in examining the properties of strange metals. Our analysis here aims to demonstrate the machine learning's capability in accurately discerning the metric \eqref{OURMET} based on holographic entanglement entropy.

%%%%%%%%%%%%%%%%%%%%%%%%%%%
\subsubsection{Linear-axion model}
As a warm-up, we verify the efficacy of our machine learning approach in scenarios where $h(z)=1$, consistent with the results obtained through Bilson's method outlined in Sec. \ref{subsec33}: the linear-axion model \eqref{LAMAC}.

Within the same dataset \eqref{LAMPARA}, we input data derived from the holographic entanglement entropy of the linear-axion model into the machine learning framework. Recall the entanglement entropy depicted in Fig. \ref{RNFIGdata}, with fitting data represented by \eqref{FITTING1} and corresponding coefficients listed in table. \ref{Table1}.

\begin{figure}
\centering
    \includegraphics[width=0.49\textwidth]{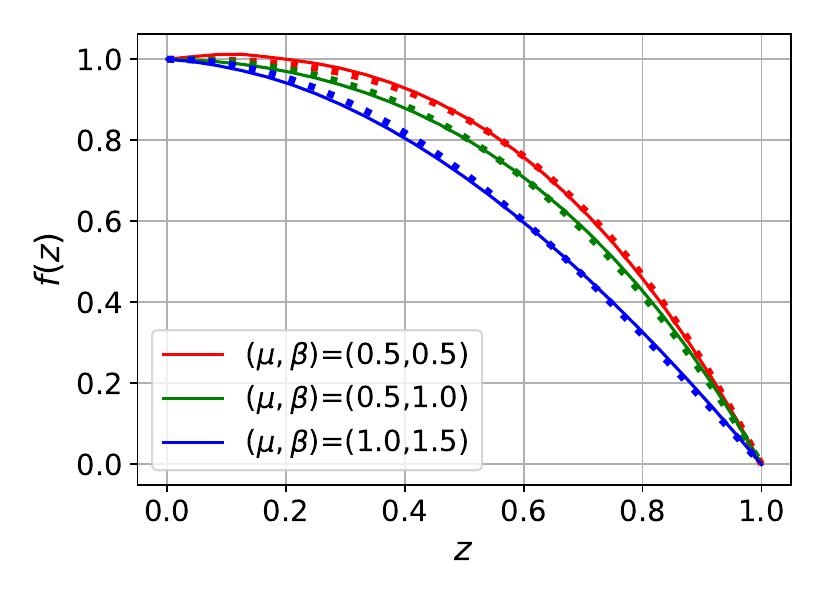}
    \includegraphics[width=0.49\textwidth]{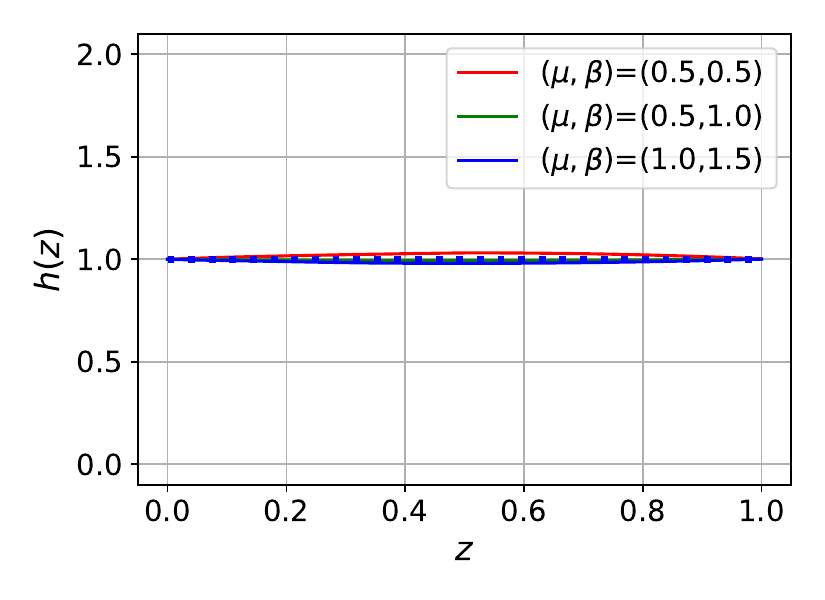}
\caption{The emergent metric from the entanglement entropy of linear-axion model with the given dataset \eqref{LAMPARA}. Solid lines are the metric from machine learning method, while dotted lines denote the true metric \eqref{LAMMET}.}\label{LMLAM}
\end{figure}

In Fig. \ref{LMLAM}, we illustrate that our machine learning methodology successfully elucidates the black hole geometry $f(z)$ within the holographic bulk spacetime. 
The emergent metric, as identified by the machine learning process, aligns closely with the geometry of the linear-axion model described by \eqref{LAMMET}. In Appendix. \ref{appb}, we discuss the detailed discussion of the training procedure employed in our machine learning method, together with the error estimation analysis.

%%%%%%%%%%%%%%%%%%%%%%%%%%%
\subsubsection{Gubser-Rocha model}
Subsequently, we apply the machine learning methodology to the Gubser-Rocha model~\cite{Gubser:2009qt} where the metric component now $h(z)\neq1$.  
Its action is defined as follows
\begin{equation}\label{GRACTION}
\begin{split}
S = \int \dd^4x\sqrt{-g}\left(  R + 6\cosh \phi - \frac{1}{4} e^\phi F^2 -\frac{3}{2}(\partial{\phi})^2  -\frac{1}{2}\sum_{I=1}^{2}(\partial \psi_{I})^2 \right) \,,
\end{split}
\end{equation}
where $16\pi G_N = L =1$. Here, $\phi$ represents the dilaton field, the field strength is $F=\dd A$, and the axion field $\psi_I$ is also included. Notably, the Gubser-Rocha model \eqref{GRACTION} stands out as one of the most prominent and celebrated holographic models, pivotal in elucidating the properties of strange metals. Remarkably, the Gubser-Rocha model facilitates the study of linear-in-$T$ resistivity, attributed to the nature of the infrared fixed point~\cite{Davison:2013txa,Gouteraux:2014hca,Anantua:2012nj,Jeong:2018tua,Jeong:2019zab,Jeong:2021wiu,Balm:2022bju,Ahn:2023ciq,Jeong:2023ynk,Ge:2023yom,Wang:2023rca}.\footnote{It is pertinent to note that the model also manifests characteristics relevant to high-$T_c$ cuprate superconductivity, such as Homes's law in high-$T_c$ superconductors~\cite{Jeong:2021wiu,Wang:2023rca}. Furthermore, investigations into the phase diagram utilizing fermionic spectral functions~\cite{Jeong:2019zab} or conductivity~\cite{Zhao:2023qms} have been undertaken. Additionally, discussions on certain limitations in describing transport anomalies, such as the Hall angle, have also been addressed~\cite{Ahn:2023ciq,Ge:2023yom}.}

The Gubser-Rocha model \eqref{GRACTION} possesses another attractive feature in that it allows for the analytic background solutions. Within our metric \eqref{OURMET}, it is described as
\begin{equation}\label{GRMET}
\begin{split}
f(z) &= (1-z) \frac{1+(1+3 Q){z} + (1+3Q(1+Q)-\frac{1}{2}\beta^{2}){z}^2}{(1+Q {z})^{3/2}} \,, \qquad  h(z) = (1+Q {z})^{3/2} \,,
\end{split}
\end{equation}
accompanied by the matter fields
\begin{equation}\label{}
\begin{split}
A &=(1-{z}) \frac{\sqrt{3Q(1+Q)\left(1-\frac{\beta^2}{2(1+Q)^2} \right)}}{1+Q {z}} \, \dd {t} \,, \\  
\phi &=\frac{1}{2} \log[1+{z}\,Q] \,, \qquad \psi_{I} = \beta (x\,,y)  \,.
\end{split}
\end{equation}
In these solutions, a free parameter, $Q$, is present. This parameter will be determined once the physical parameters, the chemical potential $\mu$ and the strength of momentum relaxation $\beta$, are given through
\begin{align}\label{}
\mu := A_t(0) = \sqrt{3Q(1+Q)\left(1-\frac{\beta^{2}}{2(1+Q)^{2}}\right)}  \,, \qquad s = 4\pi \sqrt{(1+Q)^3} \,,
\end{align}
where we also identify the thermal entropy $s$ in \eqref{TSfor} for our machine learning approach.\\

To initiate our machine learning methodology, we initially compute the holographic entanglement entropy. Since the Gubser-Rocha model entails two free parameters: ($\mu,\,\beta$), akin to the linear-axion model \eqref{LAMAC}, we opt for the same dataset \eqref{LAMPARA}.

\begin{figure}
\centering
\includegraphics[width=0.5\textwidth]{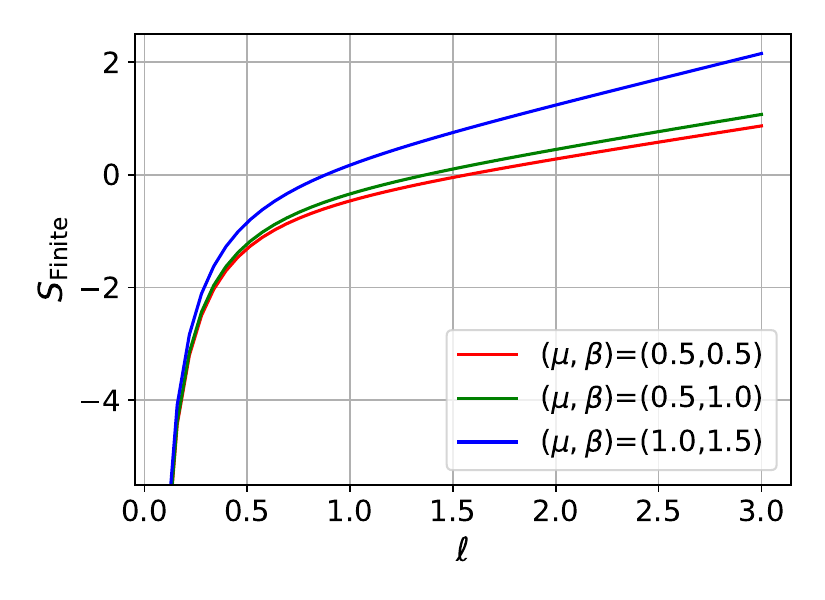}
\caption{The finite piece of the entanglement entropy of Gubser-Rocha model when $(\mu,\, \beta)=$ Data 1 (red), Data 2 (green), Data 3 (blue): see \eqref{LAMPARA}. The numerically obtained $S_{\text{Finite}}$ is aligned with the fitting \eqref{FITTING1} with the coefficients in Table. \ref{Table2}.}\label{GRFIGdata}
\end{figure}

Subsequently, by substituting \eqref{GRMET} into \eqref{SFINITE}, one can compute the entanglement entropy numerically. Employing the parameters specified in \eqref{LAMPARA}, we depict $S_{\text{Finite}}$ in Fig. \ref{GRFIGdata},
where we find its compatibility with \eqref{FITTING1} when fitted, with coefficients detailed in Table. \ref{Table2}.
\begin{table}[h]
\centering
\begin{tabular}{lrrrrrrr}
             & $c_{\,-1}$ & $c_{0}$ & $c_{1}$ & $c_{2}$  & $c_{3}$ & $c_{4}$ & $c_{5}$   \\
\hline
\hline
Data 1 & -0.71778 & 0.06566 & 0.03271 & 0.19084 & -0.04034  & 0.00441 & -0.00018   \\
Data 2 & -0.71778 & 0.09288 & 0.17478 & 0.13203 & -0.02557 & 0.00273 & -0.00012    \\
Data 3 & -0.71776 & 0.35355 & 0.39487 & 0.17682 & -0.04421 & 0.00700 & -0.00047
\end{tabular}
\caption{Numerical fitting coefficients for \eqref{FITTING1} of Gubser-Rocha model.}\label{Table2}
\end{table}

It is worth noting that the value of $c_0$ for the Gubser-Rocha model is finite, unlike that for the linear-axion model as shown in Table. \ref{Table1}. To elucidate this distinction, we analytically investigate $c_0$ for our metric \eqref{OURMET} in the limit of a small subsystems, as detailed in Appendix. \ref{appa}. Specifically, we examine $c_0$ as follows
\begin{align}\label{c0FOR}
c_0 = \frac{h'(0)}{2} \,.
\end{align}
We observe that when $h(z)=1$, $c_0$ equals 0, consistent with the numerical findings in the linear-axion model. Conversely, when $h(z)\neq1$, $c_0$ can be non-zero. This is evident in the case of the Gubser-Rocha model \eqref{GRMET}: we checked that our numerical results in Table. \ref{Table2} align with the analytical formula \eqref{c0FOR}.

\begin{figure}
\centering
    \includegraphics[width=0.49\textwidth]{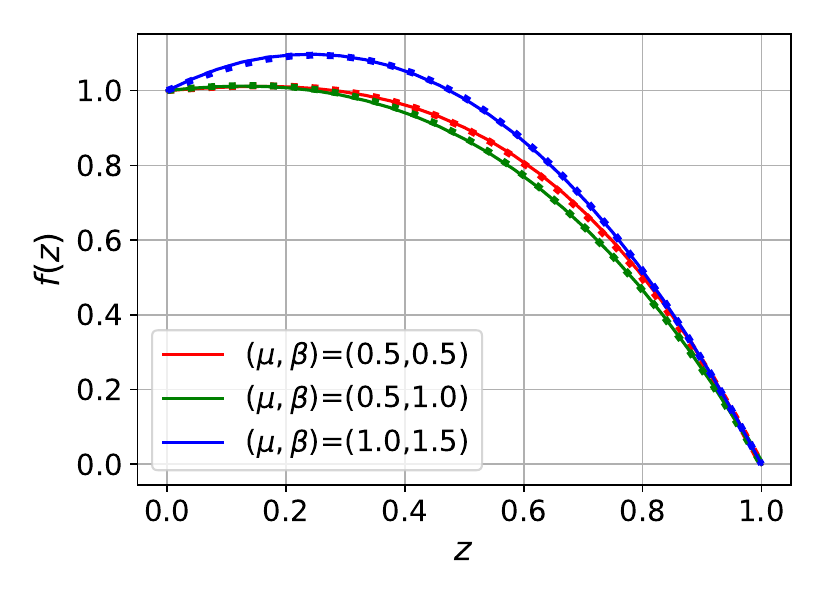}
    \includegraphics[width=0.49\textwidth]{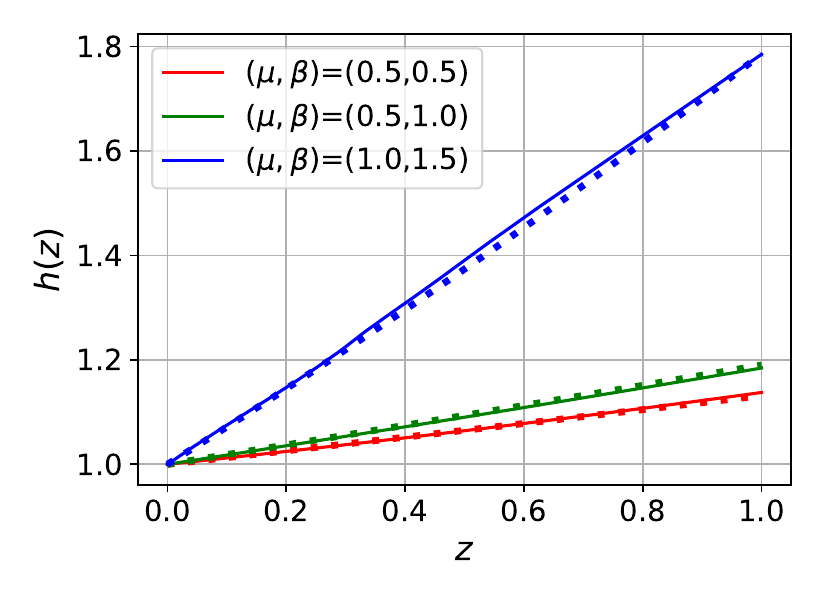}
\caption{The emergent metric from the entanglement entropy of Gubser-Rocha model with the given dataset \eqref{LAMPARA}. Solid lines are the metric from machine learning method, while dotted lines denote the true metric \eqref{GRMET}.}\label{LMGR}
\end{figure}

Finally, we plug data obtained from the holographic entanglement entropy of the Gubser-Rocha model into the machine learning framework, utilizing the fitting data described by \eqref{FITTING1} and the corresponding coefficients outlined in Table. \ref{Table2}. Subsequently, in Fig. \ref{LMGR}, we demonstrate that our machine learning methodology can reveal the complete black hole geometry $f(z)$ and $h(z)$ within the holographic bulk spacetime. This validation underscores the efficacy of our machine learning approach in capturing the entire geometry, even in scenarios where $h(z)\neq1$.
See also Appendix. \ref{appb} for the error estimation analysis of Gubser-Rocha model.

%%%%%%%%%%%%%%%%%%%%%%%%%%%
\subsubsection{Holographic superconductors}

We proceed to discuss another prominent example of spacetime within holography, characterized by $h(z)\neq1$, known as the holographic superconductor model~\cite{Hartnoll:2008vx,Hartnoll:2008kx}, often referred to as the HHH model after its proponents, Hartnoll, Herzog, and Horowitz. This model stands as one of the most widely studied frameworks in applied holography to condensed matter systems, garnering significant attention in the field (for comprehensive reviews, refer to \cite{Herzog:2009xv,Cai:2015cya}).

Precisely speaking, the U(1) symmetry in the dual field theory of the HHH model is global rather than local, characterizing it as a superfluid rather than a superconductor. While some aspects, such as electric conductivity, may not distinguish between the two, other features, such as the dynamics of vortices and collective low-energy modes, markedly differentiate between a superfluid and a superconductor. For instance, collective excitations in holographic superfluids are second sound waves (superfluid Goldstone mode)~\cite{Amado:2009ts,Arean:2021tks} which are absence in the superconductors. Explorations of collective excitations in holographic superconductors are conducted in~\cite{Jeong:2023las}, incorporating dynamical boundary gauge fields and explicitly confirming the Anderson-Higgs mechanism. Nevertheless, for the sake of consistency in our discussion, we continue to refer to the HHH model as representing holographic superconductors.

The HHH model~\cite{Hartnoll:2008vx,Hartnoll:2008kx} is given as
\begin{align}\label{HHHACTION}
\begin{split}
S  =  \int \dd^4x\sqrt{-g}\left( R  + 6 -\frac{1}{4} \,F^2   -|D\Phi|^2  - m^2 |\Phi|^2  \right) \,,
\end{split}
\end{align}
where it composes of field strength $F=\dd A$ and a complex scalar field $\Phi$ with the covariant derivative $D_\mu \Phi=  \left(  \nabla_\mu -i q A_\mu   \right)  \Phi$, which is for  superconducting phase. In this manuscript, we set $m^2=-2$ and $q=3$ as chosen in the original paper~\cite{Hartnoll:2008vx,Hartnoll:2008kx}.

To investigate the metric of holographic superconductors, we numerically solve the equations of motion derived from the action \eqref{HHHACTION}. For numerical convenience, we adopt the following ansatz for the metric
\begin{equation}\label{SCMETNUM}
\begin{split}
f(z) &= (1-z) U(z)\,, \qquad  h(z) = V(z) \,,
\end{split}
\end{equation}
where we impose the boundary condition $U(0)=V(0)=1$ to ensure the asymptotic AdS geometry. Additionally, for the matter fields, we choose
\begin{equation}\label{MFA}
\begin{split}
A =(1-{z}) a(z) \, \dd {t} \,, \qquad \Phi = z \, \eta(z)   \,,
\end{split}
\end{equation}
Notably, the boundary behavior of these matter fields is characterized by
\begin{align} \label{}
\begin{split}
A_t = \mu - \rho z + \cdots \,, \qquad
\Phi = \Phi^{(-)} \, z + \Phi^{(+)} \, z^2 + \cdots  \,.
\end{split}
\end{align}
According to the holographic dictionary, $\mu$ represents the chemical potential, $\rho$ stands for the charge density, and in the asymptotic form of $\Phi$, $\Phi^{(-)}$ denotes the source while $\Phi^{(+)}$ represents the condensate. In our chosen ansatz \eqref{MFA}, we straightforwardly derive the chemical potential $\mu$ as $\mu=a(0)$ and the source $\Phi^{(-)}$ as $\Phi^{(-)}=\eta(0)$.

Next, as the boundary condition for the superconducting phase, we set the source to be zero, $\Phi^{(-)}=0$, to describe the spontaneous symmetry breaking, expressed as
\begin{align} \label{}
\begin{split}
\Phi^{-}=0 \,, \qquad \text{and} \qquad \langle\mathcal{O}_2\rangle := \sqrt{2} \, \Phi^{+} \,,
\end{split}
\end{align}
where the factor of $\sqrt{2}$ in defining the condensate $\langle\mathcal{O}_2\rangle$ follows from \cite{Hartnoll:2008vx}, serving as a convenient normalization. Therefore, essentially, using the numerically obtained solutions, one can discern between the superconducting phase, characterized by $\Phi^{(+)} \neq 0$, and a normal phase, where $\Phi = 0$, by computing the temperatures using \eqref{TSfor}.
\begin{figure}
\centering
\includegraphics[width=0.5\textwidth]{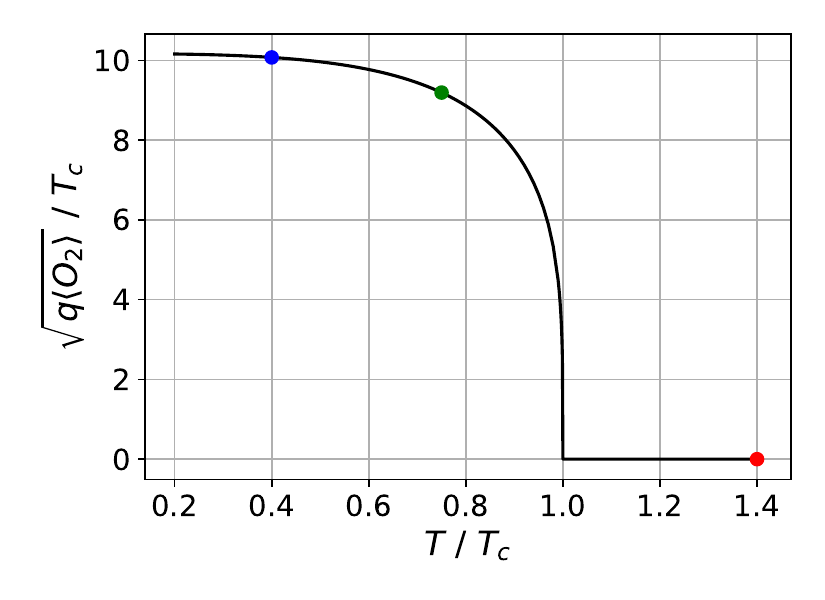}
\caption{Condensation when $m^2=-2$ and $q=3$ where the critical temperature is $T_c/\mu = 0.156$. The dataset utilized for our machine learning analysis, \eqref{DATASETFORSC}, is denoted in (red, green, blue) dots.}\label{CONDENFIG}
\end{figure}
We present the plot of the condensate in Fig. \ref{CONDENFIG}, depicting condensation when $m^2=-2$ and $q=3$, with a critical temperature of $T_c/\mu = 0.156$, consistent with findings in \cite{Hartnoll:2008kx}.\\

\begin{figure}
\centering
\includegraphics[width=0.5\textwidth]{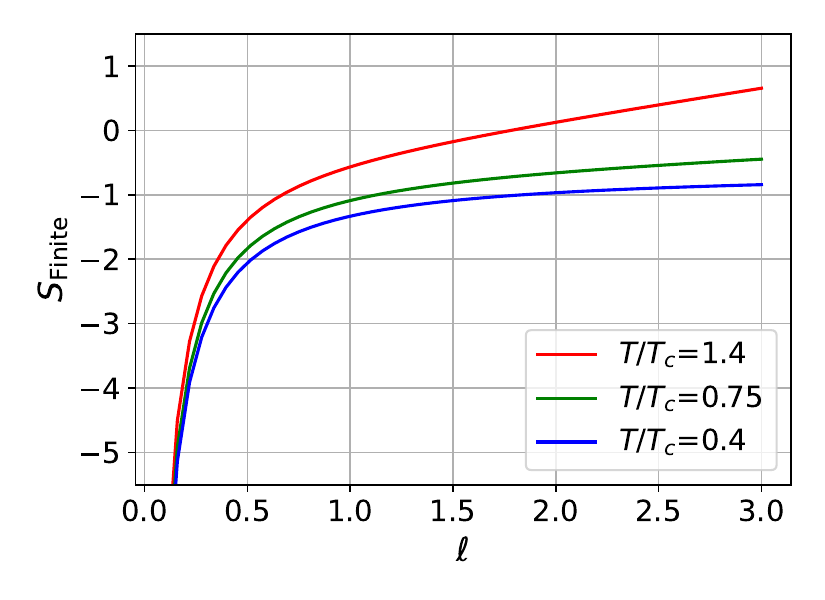}
\caption{The finite piece of the entanglement entropy of superconductor model when $T/T_c=1.4, 0.75, 0.4$ (red, green, blue). The numerically obtained $S_{\text{Finite}}$ is aligned with the fitting \eqref{FITTING1} with the coefficients in Table. \ref{Table3}.}\label{SCFIGdata}
\end{figure}

Similarly to the Gubser-Rocha model, we begin our machine learning methodology by computing the holographic entanglement entropy. For the superconductor examples, we select three representative datasets characterized by
\begin{align}\label{DATASETFORSC}
\begin{split}
T/T_c = (1.4\,,\, 0.75\,,\, 0.4) \,,
\end{split}
\end{align}
illustrated as colored dots in Fig. \ref{CONDENFIG}. Subsequently, by substituting the obtained numerical metric solutions \eqref{SCMETNUM} into \eqref{SFINITE}, we compute the entanglement entropy, as shown in Fig. \ref{SCFIGdata}.
We find its agreement with \eqref{FITTING1} when fitted, with coefficients provided in Table \ref{Table3}.
\begin{table}[h]
\centering
\begin{tabular}{lrrrrrrr}
             & $c_{\,-1}$ & $c_{0}$ & $c_{1}$ & $c_{2}$  & $c_{3}$ & $c_{4}$ & $c_{5}$   \\
\hline
\hline
$T/T_c = 1.4$ & -0.71778 & 0.00317 & -0.00981 & 0.20193 & -0.05115  & 0.00716 & -0.000397   \\
$T/T_c = 0.75$ & -0.71774 & -0.40853 & 0.01507 & 0.02504 & -0.00345 & 0.00033 & -0.000013    \\
$T/T_c = 0.4$ & -0.71771 & -0.63160 & 0.02676 & -0.01275 & 0.00304 & -0.00026 & 0.000001
\end{tabular}
\caption{Numerical fitting coefficients for \eqref{FITTING1} of superconductor model.}\label{Table3}
\end{table}
Here, we also confirmed that the analytical formula \eqref{c0FOR} yields consistent numerical values for $c_0$ in the superconductor model.

\begin{figure}
    \centering
    \includegraphics[width=0.49\textwidth]{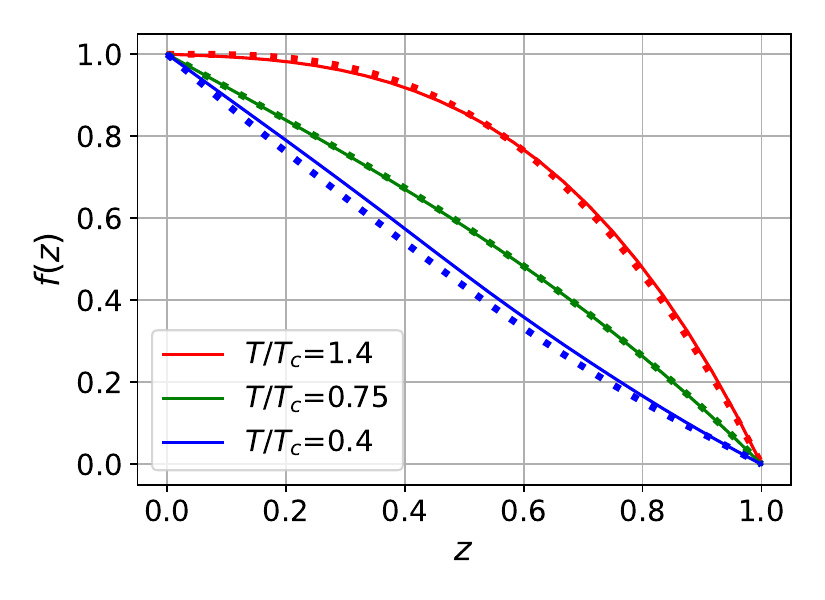}
    \includegraphics[width=0.49\textwidth]{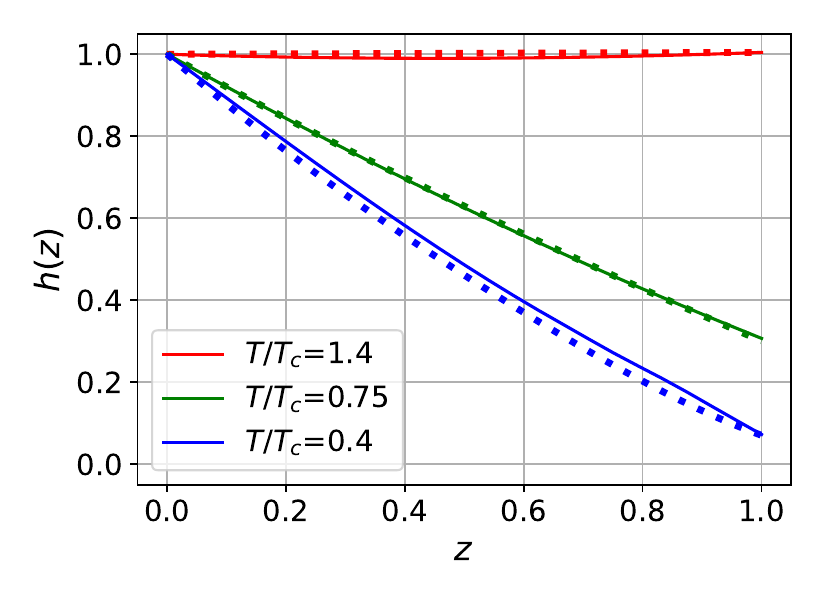}
    \caption{The emergent metric from the entanglement entropy of superconductor model. Solid lines are the metric from machine learning method, while dotted lines denote the true metric \eqref{SCMETNUM} obtained numerically solving equations of motion.}\label{LMSC}
\end{figure}

Finally, we incorporate data derived from the holographic entanglement entropy of the superconductor model into the machine learning framework. This involves utilizing the fitting data described by \eqref{FITTING1} along with the corresponding coefficients outlined in Table \ref{Table3}. Subsequently, in Fig. \ref{LMSC}, we showcase how our machine learning methodology proficiently reveals the complete black hole geometry $f(z)$ and $h(z)$ for the holographic superconductor. This validation again underscores the effectiveness of our machine learning approach in accurately capturing the entire geometry, even in scenarios where $h(z)\neq1$.
The error estimation analysis for the holographic superconductor model is given in Appendix. \ref{appb}.

It is worth noting that the functional form of $h(z)$ for superconductors differs from that of the Gubser-Rocha model (cfr. Fig. \ref{LMSC} vs. Fig. \ref{LMGR}): $h(z)$ monotonically decreases towards the horizon for superconductor models, unlike the Gubser-Rocha model. Nonetheless, we find that our machine learning approach can adeptly reconstruct all types of $h(z)$.

%%%%%%%%%%%%%%%%%%%%%%%%%%%
\subsection{Emergent spacetime from a one-dimensional chain}\label{secondim}

In this section, we demonstrate the applicability of deep learning in elucidating the emergent AdS spacetime from entanglement entropy data of many-body systems. For this purpose, we leverage data on the entanglement entropy of one-dimensional systems \cite{Miao:2021aa}, specifically focusing on data from a fermionic tight-binding chain at half filling.

It is worth noting that \cite{Miao:2021aa} have also investigated the entanglement entropy in higher-dimensional systems, which could be relevant to our AdS$_4$ setup \eqref{OURMET}. Nonetheless, unlike our holographic setup, the analysis in \cite{Miao:2021aa} does not consider strip-shaped entangling regions. Hence, in this section, we concentrate on the one-dimensional ($1+1D$) case.

%%%%%%%%%%%%%%%%%%%%%%%%%%%
\subsubsection{Machine learning setup}

\paragraph{Holographic setup.}
In order to study one-dimensional systems in holography, we consider the asymptotically AdS$_3$ spacetime as
\begin{align}\label{OURMET22}
\begin{split}
\dd s^2 = \frac{L^2}{z^2} \left[ -f(z) \, \dd t^2 +  \frac{\dd z^2}{f(z)} + h(z) \, \dd x^2 \right] \,,
\end{split}
\end{align}
where the corresponding temperature and thermal entropy read as
\begin{align}\label{BTZTDTS}
\begin{split}
T = -\frac{f'(z)}{4\pi} \Bigg|_{z_{h}} \,, \qquad s = \frac{L}{4G_N} \frac{\sqrt{h(z)}}{z} \Bigg|_{z_{h}} \,.
\end{split}
\end{align}

Then, within the metric \eqref{OURMET22}, the holographic entanglement entropy $S$ can be given \cite{Ryu:2006bv,Ryu:2006ef} as 
\begin{align}\label{ADS3FOR}
\begin{split}
S &= \frac{L}{2G_N} \left[ \int_{\epsilon}^{z_{*}} \frac{1}{z} \dd z   + \int_{0}^{z_{*}} \, \frac{1}{z} \left(\sqrt{\frac{1}{1-\frac{z^{2}}{z_{*}^{2}} \frac{h(z_{*})}{h(z)} }} \frac{1}{\sqrt{f(z)}} - 1 \right) \,\, \dd z   \right]  \\
&=  \frac{L}{2G_N} \left[  \log\left(\frac{1}{\epsilon}\right) + \left\{ \int_{0}^{z_{*}} \, \frac{1}{z} \left(\sqrt{\frac{1}{1-\frac{z^{2}}{z_{*}^{2}} \frac{h(z_{*})}{h(z)} }} \frac{1}{\sqrt{f(z)}} - 1 \right) \,\, \dd z  + \log z_{*} \right\}  \right] \\
&:= \frac{L}{2G_N} \left[  \log\left(\frac{1}{\epsilon}\right) + S_{\text{Finite}}\right] \,,
\end{split}
\end{align}
together with the subsystem size $\ell$
\begin{align}\label{ADS3FOR2}
\begin{split}
\ell = 2 \int_{0}^{z_{*}}  \frac{1}{\sqrt{\frac{h(\alpha)}{h(z_*)}\frac{z_{*}^{2}}{\alpha^{2}} - 1} } \frac{1}{\sqrt{h(\alpha)f(\alpha)}} \,\, \dd \alpha  \,.
\end{split}
\end{align}

\paragraph{Example: BTZ black holes and $1+1D$ CFT.}
One of the landmark outcomes in holographic entanglement entropy~\cite{Ryu:2006bv,Ryu:2006ef} is the utilization of the Ryu-Takayanagi formula with BTZ black holes:
\begin{align}\label{BTZMET}
\begin{split}
f(z) = 1 - z^2 \,, \qquad h(z) = 1 \,,
\end{split}
\end{align}
where $z_h=1$, yielding the consistent entanglement entropy of a $1+1D$ CFT. Substituting \eqref{BTZMET} into \eqref{ADS3FOR}-\eqref{ADS3FOR2}, one derives
\begin{align}\label{}
\begin{split}
(\text{BTZ black holes}): \quad S = \frac{L}{2 G_N} \log\left[ \frac{2}{\epsilon} \sinh\left(\frac{\ell}{2}\right)  \right] \,, \qquad \ell = 2\,\text{arctanh} (z_{*}) \,.
\end{split}
\end{align}
This aligns with the CFT result
\begin{align}\label{BTZRESULT}
\begin{split}
S_{\text{CFT}} = \frac{c}{3} \log\left[ \frac{\beta}{\pi \epsilon} \sinh\left(\frac{\pi \ell}{\beta}\right)  \right] \,,
\end{split}
\end{align}
with the identification
\begin{align}\label{BTZVALUESPARA}
\begin{split}
c = \frac{3L}{2G_N} \,, \qquad \beta = \frac{1}{T} = 2\pi \,.
\end{split}
\end{align}
where $c$ represents the central charge and $\beta$ stands for the inverse temperature derived via \eqref{BTZTDTS}.

\paragraph{Setup for machine learning method.}
Building upon the holographic setup \eqref{ADS3FOR}-\eqref{ADS3FOR2}, next we discuss our machine learning setup for one-dimensional system. Especially, to leverage the data provided in \cite{Miao:2021aa}, we introduce $S_{\text{DATA}}$ as defined in \cite{Miao:2021aa},
\begin{align}\label{SDATAFOR}
\begin{split}
S_{\text{DATA}} := S - \frac{c}{3}\log\left[ \frac{\beta v}{\pi \epsilon} \right]   \,, 
\end{split}
\end{align}
where $v$ is the group velocity, potentially varying across systems. Subsequently, comparing \eqref{SDATAFOR} with the holographic entanglement entropy $S$ in \eqref{ADS3FOR}, one can find the finite piece as
\begin{align}\label{FFORM}
\begin{split}
S_{\text{Finite}} =    \frac{3}{c} S_{\text{DATA}} + \log\left(\frac{\beta v}{\pi}\right)   \,,  
\end{split}
\end{align}
employing $c = {3L}/{(2G_N)}$. Essentially, by substituting the data $S_{\text{DATA}}$ obtained from \cite{Miao:2021aa} into the right-hand side of equation \eqref{FFORM}, we can determine the metric on the left-hand side, namely the holographic formula \eqref{ADS3FOR}.

Furthermore, in order to align with the data, we adopt the fitting curve for $S_{\text{DATA}}$ as utilized in \cite{Miao:2021aa}:
\begin{align}\label{FITFOR22}
\begin{split}
S_{\text{DATA}}  :=  \frac{c}{3} \log\left[ \sinh\left(\frac{\pi \ell}{\beta v}\right)  \right]   \,, 
\end{split}
\end{align}
where it is referred to as the universal scaling function.\footnote{In principle, $S_{\text{DATA}}$ may contain a non-universal constant, which does not affect the scaling behavior of $\ell$ and, therefore, is also not explicitly specified in \cite{Miao:2021aa}. Similarly, we omit such a term in our analysis. Notably, in the study of entanglement entropy, examining the derivative of $S$ with respect to the spatial size of the entalgment region $\ell$ can prove advantageous. This approach, as suggested by \cite{Jokela:2020auu}, offers data that are unaffected by the UV cutoff, thereby also avoiding the appearance of a non-universal constant.}
One comment is in order. By rescaling $\beta v$ to $\beta$, one may eliminate the depencence on group velocity in the formula. However, we do not adopt this approach, as it has been reported that the group velocity may vary between different systems. Specifically, the group velocity can be calculated based on the lowest energy within the symmetry sector associated with the crystal momentum. For further discussion, refer to \cite{Miao:2020hkj,Miao:2021aa}.

%%%%%%%%%%%%%%%%%%%%%%%%%%%
\subsubsection{Emergent metric from a fermionic tight-binding chain at half filling}

As a supplementary note, it is worth mentioning the following. Let us consider the entanglement entropy $S$ pertaining to the bipartition of a one-dimensional quantum many-body system into a subsystem of length $\ell$. When the system parameters and energy (temperature) fall within a quantum critical regime of the model under consideration, established principles dictate that the entanglement entropy may adhere to a universal scaling function~\cite{Shankar:1994aa,Sachdev_2011,Senthil:2008aa}.

Notably, in \cite{Miao:2021aa}, it has been demonstrated that for the typical quantum many-body systems complying with the eigenstate thermalization hypothesis, the entanglement entropy of eigenstates can be characterized by a universal scaling function. Particularly, in critical one-dimensional systems with linear dispersion at low energies, a universal scaling function follows from CFT.\footnote{The presence of gapless excitations can be indicative of a critical point. At this point, the system does not have a well-defined energy scale (gap) and exhibits critical fluctuations over all length scales. In a one-dimensional fermionic chain, the critical point manifests as a linear dispersion relation near the Fermi points $k=\pm \pi/2$ as $E(k)\approx v_F(k\mp \pi/2)$, leading to low-energy excitations where $v_F$ is the Fermi velocity.
%Thus, the gapless nature can be a hallmark of criticality, as the system can be excited with arbitrarily small energies.
}

\begin{figure}
\centering
\includegraphics[width=0.45\textwidth]{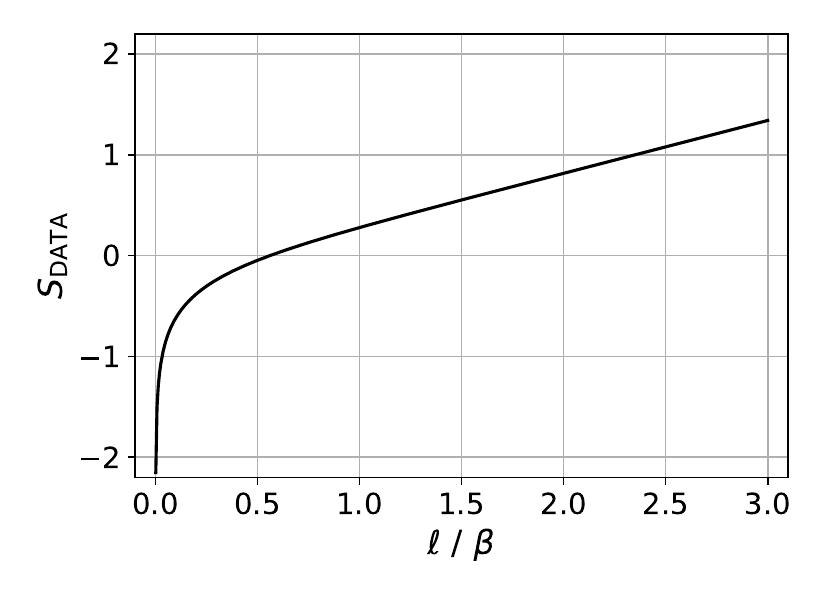}
\caption{$S_{\text{DATA}}$ for fermionic tight-binding chain at half filling \cite{Miao:2021aa}, where \eqref{FITFOR22} is used with $(c,\,v)=(1,\,2)$.}\label{FTBCHC}
\end{figure}

In this paper, within machine learning framework, we focus on analyzing data obtained from a fermionic tight-binding chain at half filling~\cite{Miao:2021aa}, where it has been demonstrated to adhere to the eigenstate thermalization hypothesis and is describable by a CFT with $c=1$.
Notably, the entanglement entropy data $S_{\text{DATA}}$ from this chain is found to be with parameters $(c,\,v)=(1,\,2)$ in \eqref{FITFOR22}: See Fig. \ref{FTBCHC}.

Furthermore, in our machine learning method, for the sake of comparison, we also investigate the BTZ case \eqref{BTZRESULT}. In essense, we have two datasets available:
\begin{align}\label{DATASETCHAIN}
\begin{split}
(c,\, v,\, \beta)=
\begin{cases}
\,\, (24\pi,\, 1,\, 2\pi)\,, \quad\,\,\, (\text{BTZ case})  \\
%%%
\,\, (1,\, 2,\, 1)\,, \qquad\quad\, (\text{Fermionic tight-binding chain at half filling})
\end{cases}
\end{split}
\end{align}
where $c = {3L}/{(2G_N)}$ with $L=16\pi G_N=1$ is employed for the BTZ case.\footnote{Combining \eqref{FFORM} with \eqref{FITFOR22}, one can notice that $c$ does not play any role in $S_{\text{Finite}}$. Nevertheless, we set $L=16\pi G_N=1$ and $\beta = 2\pi$ in order for the metric of the BTZ black hole \eqref{BTZMET}. It is noteworthy that the central charge $c$ can appear in our machine learning method through the thermal entropy \eqref{SLARGELENTROPY}.} Note that the data from \cite{Miao:2021aa}, Fig. \ref{FTBCHC}, is scaled with the inverse temperature $\beta$. This can be accommodated within the machine learning process by setting a fixed $\beta$,
\begin{align}\label{}
\begin{split}
\beta = \frac{1}{T} = -\frac{4\pi}{f'(1)} := 1\,,
\end{split}
\end{align}
where \eqref{BTZTDTS} is utilized.

Last but not least, it is worth noting that in our machine learning method aimed at determining the complete metric, we utilize information of the thermal entropy $s$ in \eqref{BTZTDTS}. When dealing with data, obtaining such thermal entropy is straightforward, as it can be derived from the slope of the entanglement entropy in the large subsystem size limit~\cite{Miao:2021aa}
\begin{align}\label{SLARGELENTROPY}
\begin{split}
(\ell \gg1): \quad S \,\approx\, S_{\text{DATA}} \,\approx\,  s \, \ell  \,,  \quad \longrightarrow \quad s = \frac{c \pi}{3 \beta v} \,, 
\end{split}
\end{align}
where we employ \eqref{FITFOR22}, which aligns with holography~\cite{Ryu:2006bv,Ryu:2006ef}.\footnote{We also confirmed that the slop of the holographic entanglement entropy in the large $\ell$ limit, given in Fig. \ref{RNFIGdata}, \ref{GRFIGdata}, \ref{SCFIGdata} are consistent with the thermal entropy from \eqref{TSfor}.}
Intuitively, when $\ell\gg1$ the subsystem effectively encompasses the entire system, implying that the minimal surface lies along the black hole horizon \cite{Hubeny:2012ry,Liu:2013una}, suggesting a potential connection between the Ryu-Takayanagi formula \eqref{EEFOR} and the thermal entropy density $s$.\\

\begin{figure}
    \centering
    \includegraphics[width=0.49\textwidth]{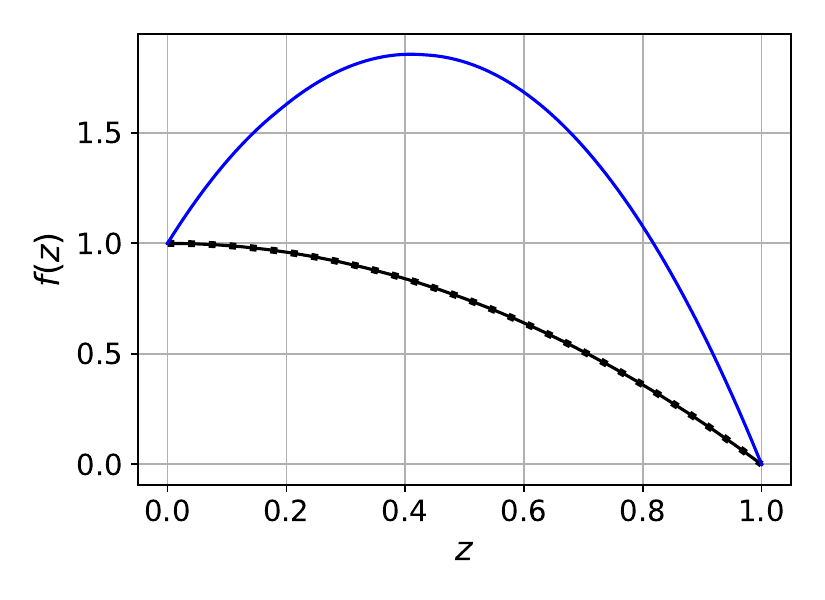}
    \includegraphics[width=0.49\textwidth]{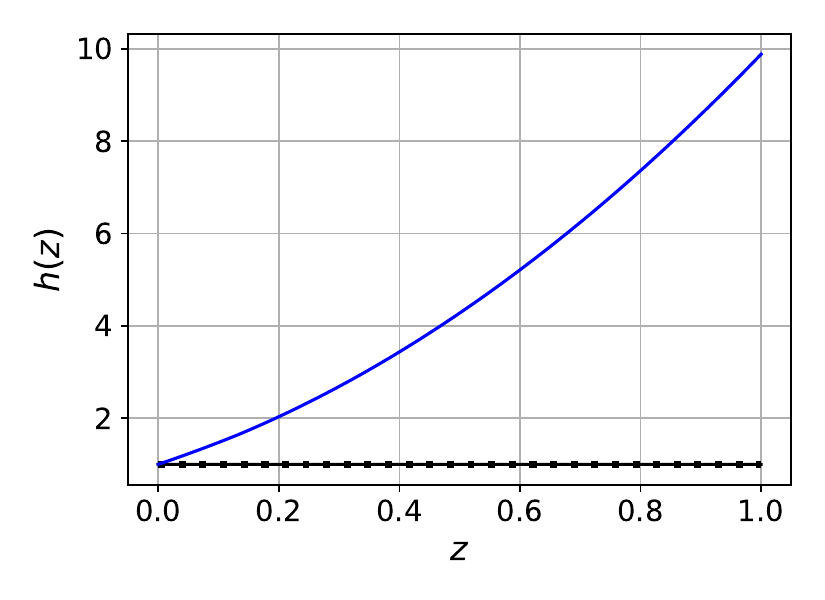}
    
\caption{The emergent metric from the entanglement entropy of BTZ black holes (black data) and fermionic tight-binding chain at half filling (blue data). Solid lines are the metric from machine learning method, while dotted lines denote the true metric of BTZ black hole \eqref{BTZMET}.}\label{FIGCHAINMACHINE}
\end{figure}

Finally, we utilize data obtained from the entanglement entropy of a fermionic tight-binding chain at half filling, as depicted in Fig. \ref{FTBCHC}, into our machine learning framework. Subsequently, as illustrated in Fig. \ref{FIGCHAINMACHINE}, we demonstrate how our machine learning methodology adeptly uncovers the complete black hole geometry $f(z)$ and $h(z)$, even leveraging entanglement entropy data from quantum many-body systems.
%

%
\iffalse
\begin{figure}
    \centering
    \includegraphics[width=0.49\textwidth]{metric_f_of_cp.pdf}
    \includegraphics[width=0.49\textwidth]{metric_h_of_cp.pdf}
\caption{The emergent metric from the entanglement entropy \eqref{FITFOR22}, utilizing the machine learning method for the parameters $(c,\,v,\,\beta)=(1,\,2,\,1)$, while varying $c'$ within $[0,0.65]$ (red-blue). The blue data corresponds to the case of the fermionic tight-binding chain at half filling in Fig. \ref{FIGCHAINMACHINE}. \HS{[HSJ: Remove this plot.]}}\label{FIGCHAINMACHINE2}
\end{figure}
\fi
%

Two noteworthy findings are made from our analysis. Firstly, as demonstrated in Fig. \ref{FIGCHAINMACHINE}, our machine learning method accurately captures the dual geometry of the CFT, specifically in the BTZ case. This validation suggests that our approach can yield continuous geometries from the entanglement entropy of the CFT, representing an advancement beyond the machine learning analysis that produces discontinuous geometries of the BTZ black holes~\cite{Park:2022fqy,Park:2023slm}.

Secondly, the functional form of the emergent metric derived from the fermionic tight-binding chain at half filling exhibits similar aspects to that of the Gubser-Rocha model: cfr. Fig. \ref{LMGR} vs. Fig. \ref{FIGCHAINMACHINE}.
The blackening factor $f(z)$ attains a maximum value, forming a peak between the AdS boundary and the horizon, while, the function $h(z)$ increases monotonically towards the horizon. This resemblance may not be surprising, as both models describe the metallic phases. In the Gubser-Rocha model, metallic behavior is characterized by the linear-temperature resistivity, whereas in the fermionic tight-binding chain, it is characterized by the gapless excitations.\\

%a non-trivial aspect in the fermionic tight-binding chain at half filling. This observation prompts consideration of employing machine learning techniques as a potential method for addressing the dual bulk geometry. Notably, the functional shape of $h(z)$ in our chain model lies intermediate to that of the Gubser-Rocha model and  superconductor models: initially showing a monotonic increase from the AdS boundary before decreasing towards the horizon.\\

We conclude this section by providing a heuristic and indicative argument for the peak structure of $f(z)$ in the fermionic tight-binding chain at half-filling. According to holographic duality, the radial coordinate $z$ in the bulk corresponds to the energy scale in the boundary theory. Thus, a peak at some $z$ may be interpreted as indicating a significant energy scale in fermionic chain. This could represent the energy at which a density of states (DOS) peak, indicating a high DOS at a particular energy level.
For instance, DOS for a one-dimentional fermionic tight-binding model is given by $D(E) \approx 1/\sqrt{4t_{h}^2 - E^2}$, where $t_{h}$ is the hopping parameter and $E$ is the energy~\cite{Ashcroft76,Kittel2004}.\footnote{Recall that the energy dispersion is $E(k)=-2t_{h} \cos k$ with $k$ being the wavevector in the first Brillouin zone, typically $k\in[-\pi,\,\pi]$ for a one-dimensional chain. Then, DOS can be computed via $D(E)=\frac{1}{2\pi}\frac{\dd k}{\dd E}$.} This form highlights the characteristic features of the DOS in such a system, indicating singularities at $E = \pm 2t_{h}$, known as Van Hove singularities.

In this context, we speculate that a peak structure in the AdS bulk, as depicted in Fig. \ref{FIGCHAINMACHINE}, may be associated with features of a fermionic tight-binding chain at half-filling, particularly as the reminiscent of high DOS due to Van Hove singularities: $E = \pm 2t_{h}$ where $t_{h}=1$ in our entanglement entropy data~\cite{Miao:2021aa}. Nevertheless, the precise nature of this peak within the bulk direction remains insufficiently understood and is likely influenced by the specific properties of the tight-binding model under consideration as well. Therefore, our argument remains speculative and a more rigorous investigation with thorough analysis are required to attain a comprehensive understanding.

%%%%%%%%%%%%%%%%%%%%%%%%%%%
%
%%%%%%%%%%%%%%%%%%%%%%%%%%%
\section{Conclusions}\label{sec5}

We have investigated the bulk reconstruction of the AdS black hole spacetime by leveraging quantum entanglement measures on the boundary field theory. Notably, by incorporating a neural network framework for the Ryu-Takayanagi formula within the AdS/DL correspondence, we introduce a machine learning approach to uncover the dual holographic emergent bulk metric from the entanglement entropy data of quantum many-body systems.

Our analysis of various entanglement entropy datasets within a  \textit{continuous} and \textit{generic} metric configuration marks significant progress in the field of machine learning holography, both practically and theoretically. By employing the neural ODE methodology together with Monte-Carlo integration, our machine learning algorithm constructs a neural network from the entanglement entropy data, where the \textit{continuous} network geometry represents the emergent holographic geometry of the quantum many-body state. Furthermore, our algorithm's ability to handle a \textit{generic} metric ansatz \eqref{OURMET} enhances its practical applicability, enabling the use of potential experimental and simulated quantum information data from strongly coupled field theories for gravity duals.

Before our work, the authors in \cite{Park:2022fqy} also applied deep learning techniques within the framework of holography to deduce the bulk metric from holographic entanglement entropy. They employed a residual neural network, composed of residual blocks, which produced a \textit{discretized} version of the holographic bulk metric. Our research has two differences from \cite{Park:2022fqy}. First, we implemented the neural ODE methodology, which utilizes a continuous function as the training variable, enabling the construction of neural networks that yield a \textit{continuous} holographic bulk metric from the entanglement entropy data. Second, our machine learning framework is capable of generating a general metric, whereas \cite{Park:2022fqy} dealt with a simplified metric ansatz ($h(z)=1$), where Bilson's method can be advantageous. Accordingly, our method may be applied to more practical and realistic gravity models of strongly coupled matter.

Additionally, the authors in \cite{Park:2023slm} explored holographic entanglement entropy using a machine learning framework, albeit with a different approach. They employed the transformer algorithm, commonly used in natural language processing, to enable the machine to learn patterns from large datasets and predict new inputs based on trained network. Nevertheless, unlike \cite{Park:2022fqy} and our approach, which focus on generating the discrete/continuous bulk metric from the entanglement entropy data, the approach in \cite{Park:2023slm} is centered on identifying patterns in the relationships between a known metric and the corresponding holographic entanglement entropy, effectively associating the constructed network with the Ryu-Takayanagi formula itself.\\

In our study, we initially utilized holographic entanglement entropy as input data for our machine learning approach. This enabled us to accurately derive the bulk metrics for both the Gubser-Rocha model and holographic superconductor models, which are prominent holographic models of strongly coupled quantum systems characterized by a non-trivial metric where $h(z)\neq1$: see Fig. \ref{LMGR} for the case of Gubser-Rocha model and Fig. \ref{LMSC} for the superconductor model.

Additionally, our machine learning approach successfully incorporates entanglement entropy data from a fermionic tight-binding chain at half filling in critical one-dimensional systems~\cite{Miao:2021aa}, and identifies the corresponding emergent bulk metric: see Fig. \ref{FIGCHAINMACHINE}. Remarkably, we observed that the emergent metric exhibits a resemblance to that of the Gubser-Rocha model, where $h(z)$ increases monotonically towards the horizon and $f(z)$ reveals an anomalous peak structure. This resemblance may be attributed to the fact that both models can describe metallic phases, with the Gubser-Rocha model featuring linear-temperature resistivity and the fermionic tight-binding chain exhibiting gapless excitations, where electrons can move freely without encountering an energy gap.

Our successful results indicate that, on a practical level, our algorithm enhances the efficiency of modeling the emergent metric from the entanglement structure of quantum many-body systems. Furthermore, it signifies the successful completion of the bulk reconstruction program, thereby advancing our understanding of holographic emergent gravity derived from quantum entanglement.\\

It is noteworthy that much of the work on metric reconstruction, including our own, has been motivated by the idea that spacetime is constructed through entanglement entropy. However, it has been argued that ``entanglement is not enough" to fully encode spacetime~\cite{Susskind:2014moa}, as the Ryu-Takayanagi surface in the gravity context only probes the exterior metrics of black holes. To gain insight into the full geometry even interior of a black hole, it has been proposed that quantum computational complexity plays a crucial role. Indeed, based on various holographic complexity proposals, such as complexity=volume~\cite{Susskind:2014moa, Susskind:2014rva}, complexity=volume 2.0~\cite{Couch:2016exn}, and complexity=anything~\cite{Belin:2021bga}, the internal metrics of black holes have been successfully reconstructed without employing machine learning techniques~\cite{Hashimoto:2021umd,Xu:2023eof}.

In general, understanding the internal structure of a black hole poses a captivating and fundamental challenge, both theoretically and experimentally. A recent and actively studied direction in holography involves examining the structure and dynamics of the black hole interior, particularly near the singularity where spacetime curvature becomes infinitely intense. For instance, pioneering work on the Kasner singularity in holography can be found in~\cite{Frenkel:2020ysx}.

As such, exploring and reconstructing the interior geometry of a black hole using machine learning holographic techniques is highly significant, not only from the perspective of quantum information theory via complexity but also for advancing the study of quantum gravity. We intend to pursue this line of research in future studies and plan to address it in subsequent investigations.

%%%%%%%%%%%%%%%%%%%%%%%%%%%%%%%%
%
%%%%%%%%%%%%%%%%%%%%%%%%%%%%%%%%
\acknowledgments
We would like to thank {Yongjun Ahn, Juan F. Pedraza} for valuable discussions and correspondence.
This work was supported by the Basic Science Research Program through the National Research Foundation of Korea (NRF) funded by the Ministry of Science, ICT $\&$ Future Planning (NRF-2021R1A2C1006791) and the AI-based GIST Research Scientist Project grant funded by the GIST in 2024.
This work was also supported by Creation of the Quantum Information Science R$\&$D Ecosystem (Grant No. 2022M3H3A106307411)
through the National Research Foundation of Korea (NRF) funded by the Korean government (Ministry of Science and ICT).
The work was also supported by the National Research Foundation of Korea (NRF) grant funded by the Korea Government (MSIT) (NRF-2023K2A9A1A01095488).
Hospitality at APCTP during the program ``Holography 2024: Correlation and Entanglement in Quantum matter'' is kindly acknowledged.
H.-S Jeong acknowledges the support of the Spanish MINECO ``Centro de Excelencia Severo Ochoa'' Programme under grant SEV-2012-0249. This work is supported through the grants CEX2020-001007-S and PID2021-123017NB-I00, funded by MCIN/AEI/10.13039/501100011033 and by ERDF A way of making Europe.
B. Ahn was supported by Basic Science Research Program through the National Research Foundation of Korea funded by the Ministry of Education (NRF-2020R1A6A3A01095962, NRF-2022R1I1A1A01064342).
All the authors contributed equally to this paper and should be considered as co-first authors.

%%%%%%%%%%%%%%%%%%%%%%%%%%%%%
%%%%%%%%%%%%%%%%%%%%%%%%%%%%%
\appendix
\section{Entanglement entropy in the small subsystem size limit}\label{appa}

In this section, we derive the holographic entanglement entropy \eqref{SFINITE} in the limit of small subsystem size $\ell\ll1$ (or small $z_{*}$). In this regime, the Ryu-Takayanagi surface is located near the AdS boundary. Consequently, we consider the asymptotic AdS behavior of the metric \eqref{OURMET} as follows
\begin{align}\label{apf1}
\begin{split}
f(z) = 1 + f'(0) \, z + \cdots \,, \qquad h(z) = 1 + h'(0) \, z + \cdots \,,
\end{split}
\end{align}
where $f'(0)=h'(0)$ is the model-dependent constant.
By substituting this expression for $\ell$ into \eqref{SLfor} and expanding the integrand for small $z_{*}$, we find
\begin{align}\label{apf2}
\begin{split}
\ell = z_{*}\frac{2\sqrt{\pi}\,\Gamma\left(\frac{7}{4}\right)}{3 \, \Gamma\left(\frac{5}{4}\right)}  \,,
\end{split}
\end{align}
which holds for small $\ell$. Similarly, utilizing \eqref{apf1} and \eqref{apf2}, we can analytically compute the holographic entanglement entropy \eqref{SFINITE} in the small $\ell$ limit as
\begin{align}\label{cofor}
\begin{split}
S_{\text{Finite}} =  - \frac{2\pi}{\ell} \left(\frac{\Gamma\left(\frac{3}{4}\right)}{\Gamma\left(\frac{1}{4}\right)}\right)^2 + \frac{h'(0)}{2}  + \cdots \,,
\end{split}
\end{align}
where $\cdots$ denotes the $\mathcal{O}(\ell)$ correction, which can be obtained by considering the $z^2$ order in the metric in \eqref{apf1}. Comparing \eqref{FITTING1} with \eqref{cofor}, we can determine $c_0$ as \eqref{c0FOR}.

%%%%%%%%%%%%%%%%%%%%%%%%%%%%%
\section{The procedure of training the network and the error estimation}\label{appb}

In this section, we discuss the training procedure employed in our machine learning approach, along with the error estimation.

\begin{figure}[]
\centering
\subfigure[Linear-axion model]
{\includegraphics[width=1.0\textwidth]{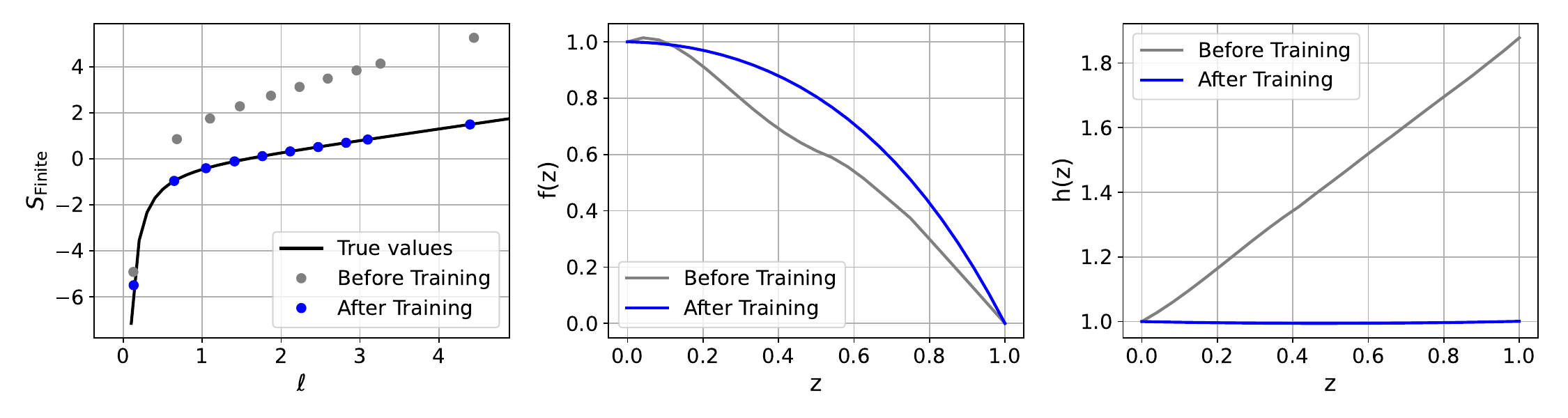}}
\subfigure[Gubser-Rocha model]
{\includegraphics[width=1.0\textwidth]{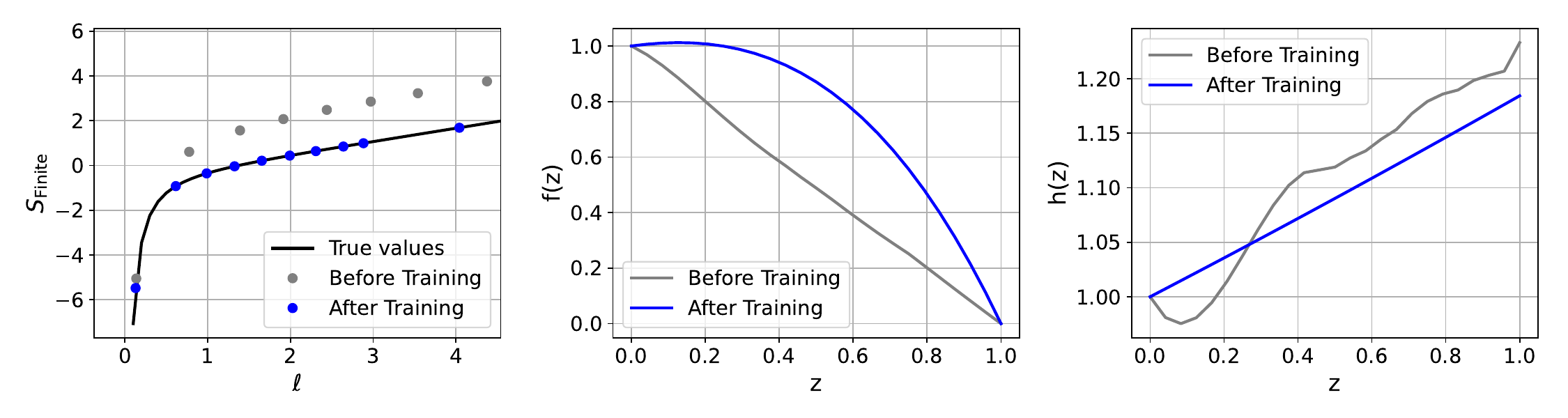}}
\subfigure[Holographic superconductor model]
{\includegraphics[width=1.0\textwidth]{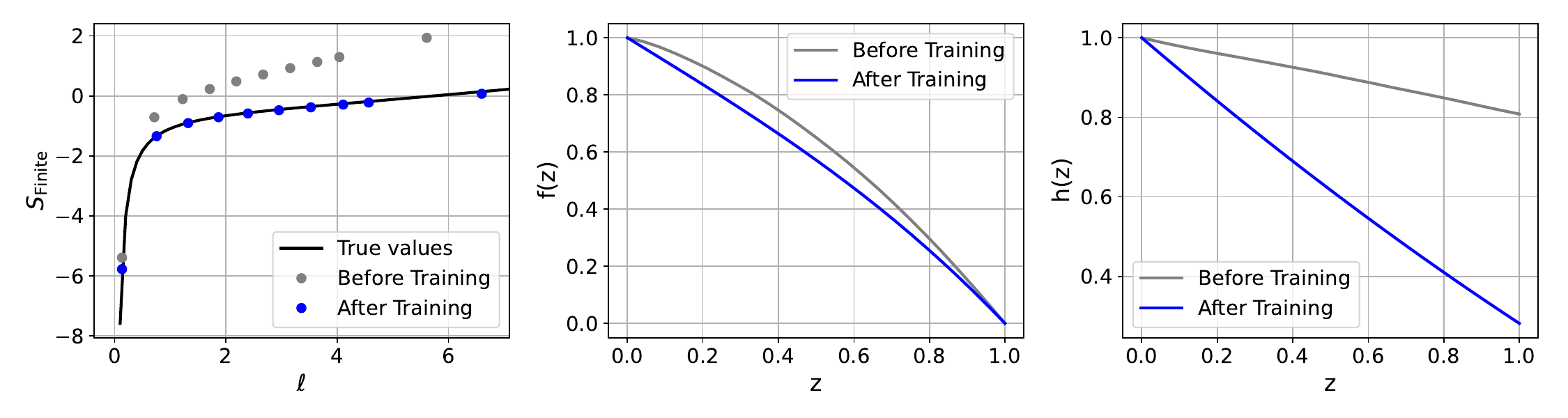}}
\subfigure[Tight-binding chain model]
{\includegraphics[width=1.0\textwidth]{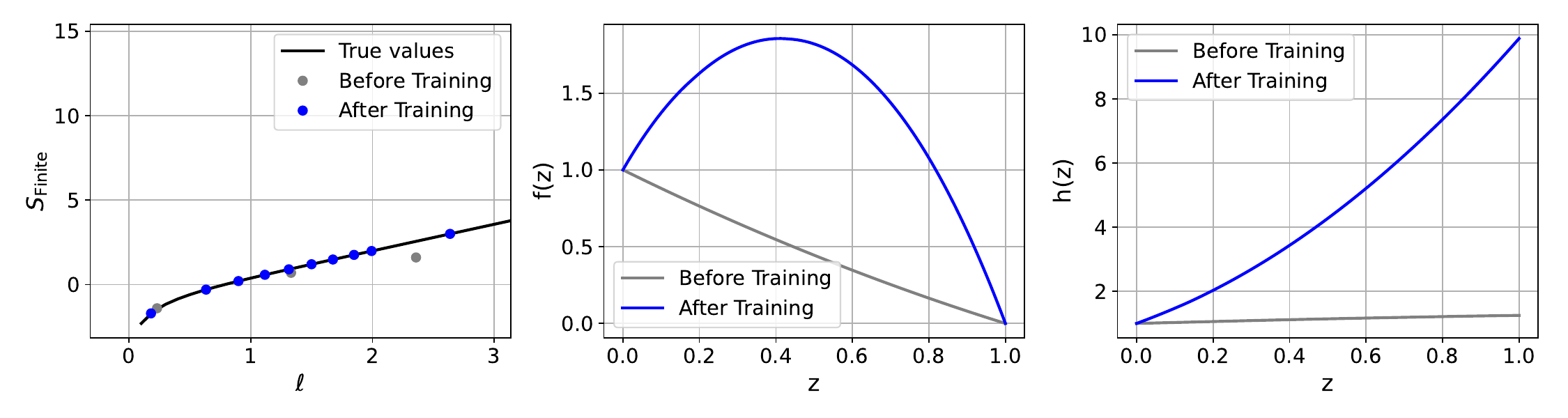}}
\caption{Machine learning data before and after training. For the holographic models (a)-(b), Data 2 from \eqref{LAMPARA} is utilized, and $T/T_c=0.75$ is for the model (c). In all figures, the gray data represents the results before training, whereas the blue data indicates the results after training. The black solid line in $S_{\text{Finite}}$ represents the input entanglement entropy data.}\label{AppxBfig}
\end{figure}

In Fig. \ref{AppxBfig} (a)-(c), we illustrate the representative results for the holographic models, namely, the linear-axion model, the Gubser-Rocha model, and the holographic superconductor model, using Data 2 from \eqref{LAMPARA} and $T/T_c=0.75$.
The figure clearly demonstrates that throughout the training process, the metric functions $f(z)$ and $h(z)$ gradually converge towards the values (indicated by blue lines) obtained from the holographic models: the gray lines represent the metric components initialized (i.e., before training) with random values on weights $W^{ij}$ and biases $b^i$ of the deep neural networks. Additionally, in Fig. \ref{AppxBfig} (d), we depict the training procedure for the tight-binding chain model.

In the figure, we also present both the untrained and trained entanglement entropy data for all the models under consideration. %This can serve as an error estimation via the loss function \eqref{LossFUNC}. 
It is evident that the entanglement entropy calculated with the trained metrics, represented by blue dots, closely aligns with the input entanglement entropy data (depicted by the black solid line).

\begin{figure}
\centering
\includegraphics[width=0.8\textwidth]{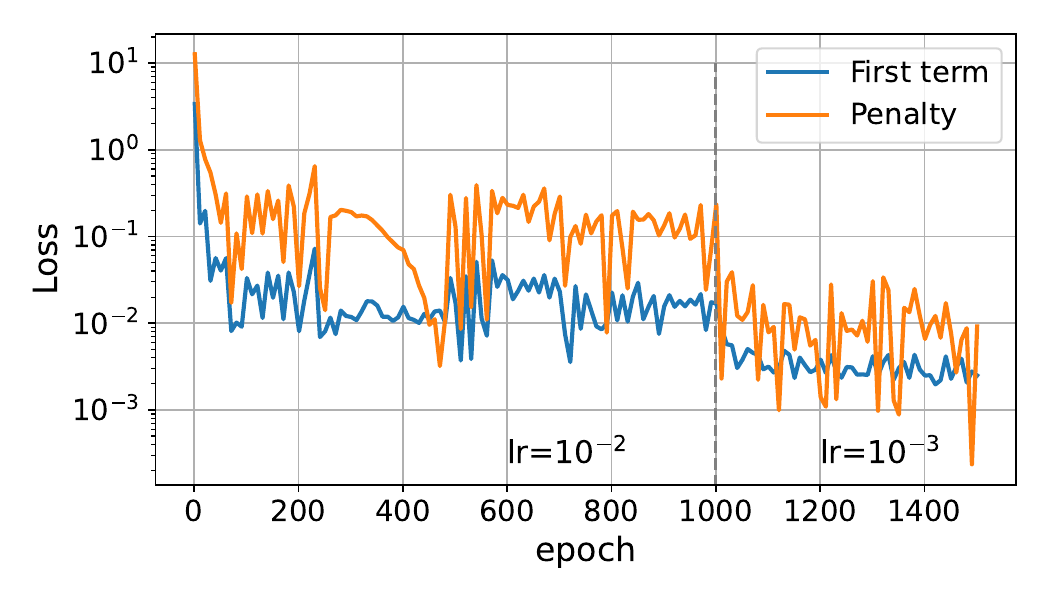}
\caption{The first and second term of the Loss function \eqref{LossFUNC} of Data 2 in the Gubser-Rocha model. The horizontal axis (epoch) denotes the number of updating the training parameters. The learning rate (lr) is set up by $10^{-2}$ before the 1000 epoch, and by $10^{-3}$ after that.}\label{Appx5}
\end{figure}

We also evaluate the loss function \eqref{LossFUNC} estimated from our machine learning method. 
For all datasets, we find that the values of both the first and second terms in \eqref{LossFUNC} consistently remain within the range of $10^{-4}$ to $10^{-2}$ after training: for instance, see Fig. \ref{Appx5} for the Gubser-Rocha model.
%For all datasets, we find that the value of the first term in \eqref{LossFUNC} remains on the order of $10^{-3}$ after training, while the second term (penalty term) is approximately $10^{-2}$: for instance, see Fig. \ref{Appx5} for the Gubser-Rocha model.
It is expected that as these numerical values decrease further, for instance to the order of $10^{-10}$, the error in our machine learning results (e.g., a small deviation in the blue data in Fig. \ref{LMSC}) will correspondingly diminish.


\clearpage

%%%%%%%%%%%%%%%%%%%%%%%%%%%%%%%%%%%%%%%%%%%%%%%%%%%
% 
%%%%%%%%%%%%%%%%%%%%%%%%%%%%%%%%%%%%%%%%%%%%%%%%%%%

%\bibliography{Refs}
\bibliographystyle{JHEP}
\providecommand{\href}[2]{#2}\begingroup\raggedright\endgroup
\end{document}